\documentclass[12pt]{iopart}

\usepackage{epsfig}

\usepackage{iopams}

\begin{document}

\title[A cosmological model in Weyl-Cartan spacetime II]{A cosmological model
  in Weyl-Cartan spacetime: \quad
  II. Magnitude-redshift relation}
\author{Dirk Puetzfeld}
\address{Institute for Theoretical Physics, University of Cologne,
  50923 K\"oln, Germany}
\ead{dp@thp.uni-koeln.de}

\begin{abstract}
In this second part of our series of articles on alternative cosmological
models we investigate the observational consequences for the new
Weyl-Cartan model proposed earlier. We review the derivation of the
magnitude-redshift relation within the standard FLRW model and characterize 
its dependence on the underlying cosmological model. With this knowledge at
hand we derive the magnitude-redshift relation within our new Weyl-Cartan
model. We search for the best-fit parameters by using the combined data set of
92 SNe of type Ia as compiled by Wang, which is based on recent supernova data
of Perlmutter \etal and Riess \etal. Additionally, we compare our best-fit
parameters with the results of several other groups which performed similar
analysis within the standard cosmological model as well as in non-standard models.   
\end{abstract}

\submitto{\CQG}
\maketitle

\section{Introduction\label{Introduction_section}}

This article represents the second part of our series on cosmological models 
within alternative gravity theories. In the first part \cite{Extweyl}, we 
presented a cosmological model in Weyl-Cartan spacetime which is no longer 
tied to a Riemannian spacetime and is based on an
alternative gravity theory, so-called metric-affine gravity (MAG), which was
essentially developed by Hehl \etal and has been extensively reviewed in
\cite{PhysRep}. In \cite{Extweyl} we derived the field equations of our new
model and provided exact solutions for a rather broad class of parameters.
We were able to show that it is possible to construct models in which the
non-Riemannian quantities die out with time. 

Within this work we investigate the observational consequences of the new model,
and try to rule out if it is compatible with current observations. As becomes 
clear from the title we are interested in the
derivation of the so-called magnitude-redshift relation, which establishes a 
relation between the
luminosity and distance of a given object and the parameters within the
underlying cosmological model. This relation was successfully used to
extract parameters within the cosmological standard model, also known as 
Friedman-Lema\^{\i}tre-Robertson-Walker (FLRW) model \cite{KolbTurner,Peacock}, 
which is based on General Relativity (GR). 

Recent observations \cite{Perlmutter,Perlmutter2,Schmidt,Riess1,Riess2} of type 
Ia supernovae, which are believed to be some kind of {\it standard candles}, led 
to the unexpected discovery that we seem to live in an universe which is currently
undergoing an accelerated phase of expansion \cite{Schmidt,Riess2}. The availability 
of a data set incorporating over 90 type Ia supernovae has turned the
magnitude-redshift relation into a cosmological test which must be passed by
any new model. It interesting to test if predictions, like the one that the expansion 
of the universe is currently accelerating, still hold in case of our Weyl-Cartan model. 
Therefore we put special emphasis on the model dependence of certain assumptions which 
are made during the derivation of the magnitude-redshift relation. It is noteworthy 
that the analysis of the SNe data led to a renewed interest in the cosmological constant $\lambda$ within the cosmological standard model. Since the cosmological constant seems to be a concept which re-emerges periodically within cosmology it is interesting to figure out if it is also inevitable for the description of the data within an alternative model. Note that there are also several efforts, cf.\ \cite{Moertsell,Csaki,Deffayet} e.g., to cope with the new observational situation, which do not focus on a change of the underlying cosmological model.

Although parts of this work are based on the results obtained in
\cite{Extweyl}, it should also be of use for the reader who wants to learn
more about the derivation and the use of the magnitude-redshift relation
within the standard FLRW model. At this point we like to mention that the
model in Weyl-Cartan spacetime still has toy model character, since many 
important issues are not yet worked out. Hence this article represents a first
attempt to rule out whether our new model is viable when confronted with the
currently available observational data from the type Ia supernovae.

The plan of the paper is as follows. In section \ref{Mag_redshit_in_standard_model} 
we review some standard concepts of the FLRW model and introduce the notions and 
assumptions which are necessary to derive the magnitude-redshift relation within 
this model. In section \ref{EXTWEYL_magnitude_redshift_relation} we make use of 
these results in order to derive this relation within the new Weyl-Cartan model. 
We perform fits within the FLRW as well as within the Weyl-Cartan model to the combined 
data set of Wang \cite{Wang} in section \ref{Numerical_results_section}. Finally, 
we compare our results with the results of several other groups who performed
similar analysis within the FLRW as well as in non-standard model and draw our
conclusion in section \ref{CONCLUSION_section}. In \ref{NATURAL_UNITS}, we provide 
an overview over the units used throughout the preceding sections.

\section{Magnitude-redshift relation within the cosmological standard model\label{Mag_redshit_in_standard_model}}

\bigskip In this section we present a short derivation of the
magnitude-redshift relation within the standard FLRW model. The results
obtained here will be of use when we perform our analysis within the new
Weyl-Cartan model in the next section. In particular we will stress the
model dependence of the magnitude-redshift relation in the following
section, especially its entanglement with the field equations. Since the
upcoming derivation of the field equations can be found in every standard
textbook on cosmology we keep ourselves rather short. For further reading the 
reader is referred to \cite{KolbTurner,Peacock}.

\subsection{Field equations\label{FLRW_field_equations_kapitel}}

The assumption of homogeneity and isotropy leads to the so-called Robertson-Walker 
metric as starting point. Using spherical coordinates $(t,r,\theta ,\phi )$ and 
the coframe 
\begin{equation}
\vartheta ^{\hat{0}}=dt,\quad \vartheta ^{\hat{1}}=\frac{S(t)}{\sqrt{1-kr^{2}}}\,dr,\quad \vartheta ^{\hat{2}}=S(t)\,r\,d\theta ,\quad \vartheta ^{\hat{3}}=S(t)\,r\,\sin \theta \,d\phi ,  \label{spherical_coframe}
\end{equation}
the line element is given by 
\begin{equation}
ds^{2}=-\vartheta ^{\hat{0}}\otimes \vartheta ^{\hat{0}}+\vartheta ^{\hat{1}}\otimes \vartheta ^{\hat{1}}+\vartheta ^{\hat{2}}\otimes \vartheta ^{\hat{2}}+\vartheta ^{\hat{3}}\otimes \vartheta ^{\hat{3}}.
\label{FLRW_robertson_walker_line_element}
\end{equation}
The function $S(t)$ is the cosmic scale factor, and $k$ can be chosen to be $+1$, $-1$, or $0$ for spaces of constant positive, negative, or zero spatial curvature, respectively. The only thing missing in order to set up the field equations is an appropriate matter model. Following the cosmological standard model, we assume that matter is smoothly smeared out over the whole spacetime. Thus, we choose the energy-momentum 3-form of an ideal fluid with pressure $p$, energy-density $\mu $, and four-velocity $u^{\alpha }$, \thinspace i.e.\ 
\begin{equation}
\Sigma _{\alpha }^{\texttt{\tiny \rm \rm fluid}}=\Sigma _{\alpha \beta }\,\eta
^{\beta }, \texttt{\small \rm \quad }  \label{fluid_energy_3_form}
\end{equation}
where $\eta^\alpha:={}^{\star}\vartheta^{\alpha}$ and
\begin{equation}
\Sigma ^{\alpha \beta }=\left( \mu +p\right) u^{\alpha }u^{\beta}+pg^{\alpha \beta }.  \label{classical_ideal_fluid_cosmology}
\end{equation}
The Einstein field equations with cosmological constant read: 
\begin{equation}
\eta _{\alpha \beta \gamma }\wedge \tilde{R}^{\beta \gamma }+2\lambda \eta
_{\alpha }=2\kappa \Sigma _{\alpha }^{\texttt{\tiny \rm \rm fluid}}.\label{field_eq_fluid_classic}
\end{equation}
Here $\tilde{R}_{\alpha \beta }$ denotes the Riemannian curvature 2-form, $\eta^{\alpha \beta \gamma}:={}^{\star}\left({\vartheta^{\alpha} \wedge
    \vartheta^{\beta} \wedge \vartheta^{\gamma}}\right)$, $\lambda$ is the
    cosmological constant, and $\kappa$ is the gravitational coupling constant. Insertion of the
Roberston-Walker metric yields the following set of field equations (the dot
denotes differentiation with respect to $t$) 
\begin{eqnarray}
\left( \frac{\dot{S}}{S}\right) ^{2}+\frac{k}{S^{2}}-\frac{\lambda }{3} &=&\frac{\kappa }{3}\mu ,  \label{friedman1} \\
2\frac{\ddot{S}}{S}+\left( \frac{\dot{S}}{S}\right) ^{2}+\frac{k}{S^{2}}-\lambda  &=&-\kappa p.  \label{friedman2}
\end{eqnarray}
Hence the field equations (\ref{field_eq_fluid_classic}) turned into a set of
ordinary differential equations for the scale factor $S(t)$. The set of
additional unknown parameters $\lambda $, $\,k$, $\mu $, and $p$ depends on
the model we decide to consider. Note that $\mu $ and $p$ are related by an
equation of state (eos) $p=p(\mu )$.

In addition to the field equations (\ref{friedman1})-(\ref{friedman2}) we have one Noether identity, which takes the following form in a Riemannian spacetime 
\begin{equation}
D\Sigma _{\alpha }=0.  \label{FLRW_first_noehter_riemannian}
\end{equation}
Let us assume that the equation of state takes the form $p\left( t\right) =w$ $\mu \left( t\right) $, with $w$ $=\,$const. Using (\ref{spherical_coframe})--(\ref{FLRW_robertson_walker_line_element}), and (\ref{fluid_energy_3_form}), equation (\ref{FLRW_first_noehter_riemannian}) turns into 
\begin{equation}
\dot{\mu}S=-3\dot{S}\,\left( \mu +p\right) \quad \stackrel{p = w
  \mu}{\Rightarrow} \quad \mu =\varkappa _{1}S^{-3\left( 1+w\right) }.
\label{FLRW_energy_density_scale_factor_relation_for_arbitrary_w_const}
\end{equation}
Thus, we have found a relation between the energy density and the scale
factor, which depends on the constant $w$ in the equation of state and an
integration constant $\varkappa_{1}$.  

\subsection{Critical density\label{FLRW_critical_density_kapitel}}

In case of a flat universe, $k=0$, the Hubble function $H:=\dot{S}/S$ and the density $\mu $ are related via a unique function $\mu _{c}=\mu (H)$,
which is often called the \textit{critical density}. The critical density is
obtained via the first Friedman equation (\ref{friedman1}), in case of a
vanishing cosmological constant we have\footnote{Note that we make use of
  natural units (cf.\ \ref{NATURAL_UNITS}), i.e.\ $\hbar =c=1$. Thus, the
  gravitational coupling constant becomes $\kappa = 8\pi G$.} 
\begin{equation}
H^{2}+\frac{k}{S^{2}}=\frac{\kappa }{3}\mu =\frac{8\pi G}{3}\mu \quad \stackrel{k=0}{\Rightarrow} \quad \mu _{c}:=\frac{3H^{2}}{8\pi G}.
\label{critical_density_in_FLRW_kapitel}
\end{equation}
This quantity is called critical density because of its character to distinguish between a open, flat, or closed universe, i.e.\ 
\begin{equation}
\mu =\left( H^{2}+\frac{k}{S^{2}}\right) \frac{3}{8\pi G}\quad \Rightarrow
\quad \mu _{k=-1}<\mu _{c}<\mu _{k=1}.
\end{equation}
As we can infer from (\ref{critical_density_in_FLRW_kapitel}) the critical density is determined by measuring the
current value of the Hubble function. Let us introduce the dimensionless 
\textit{density parameter} $\Omega_w$, which represents the relationship
between the actual and the critical density 
\begin{equation}
\Omega_w =\frac{\mu }{\mu _{c}}=\frac{8\pi G}{3H^{2}}\mu =\frac{\kappa }{3H^{2}}\mu.
\end{equation}
Note that we use $w$ as an index since we did not specify the underlying
equation of state. In case of a non-vanishing cosmological constant $\lambda$ the first Friedman
equation might be written in terms of a total density parameter which
encompasses both contributions, i.e.\ $\Omega _{\texttt{\tiny  \rm total}}=\Omega_w +\Omega _{\lambda}:=\frac{\kappa }{3H^{2}}\mu+\frac{\lambda }{3H^{2}}$, yielding
\begin{equation}
H^{2}+\frac{k}{S^{2}}-\frac{\lambda }{3}=\frac{\kappa }{3}\mu \quad
\Leftrightarrow \quad \Omega _{\texttt{\tiny \rm \rm total}}-1=\frac{k}{S^{2}H^{2}}.
\label{FLRW_first_friedman_written_with_omega_total}
\end{equation}
Note that the value of total density parameter distinguishes between the
three possible geometries of the 3-dimensional subspace, i.e.\ 
\begin{equation}
\quad \Omega _{\texttt{\tiny \rm \rm total}}\left\{ 
\begin{tabular}{l}
$<1$ \\ 
$=1$ \\ 
$>1$
\end{tabular}
\right. \Rightarrow k\left\{ 
\begin{tabular}{ll}
$<0$ & open \\ 
$=0$ & flat \\ 
$>0$ & closed
\end{tabular}
\right. .  \label{FLRW_omega_total_geometry_realtion}
\end{equation}

\subsection{Redshift\label{FLRW_redshift_subsection}}

Maybe one of the most striking properties of the cosmological standard model is the redshift due to the global expansion. In the following we will focus on
its derivation by using our ansatz for the metric from equations (\ref{spherical_coframe})--(\ref{FLRW_robertson_walker_line_element}).

The first assumption we make is that the propagation of light coming from a
distant object, a galaxy e.g., can be treated as a classical wave
phenomenon. With $r=r_{1}$ being the radial coordinate of the object, $r=r_{0}$ coordinate of the observer, $t=t_{1}$ time at which wave is emitted, and $t=t_{0}$ at which the light is detected at $r=r_{0}$, the Robertson-Walker line element yields 
\begin{equation}
ds^{2}=0 \stackrel{\theta=\phi=\rm{const}}{\Rightarrow}
\int_{t_{1}}^{t_{0}}\frac{dt}{S(t)}=\int_{r_{0}}^{r_{1}}\frac{dr}{\sqrt{1-kr^{2}}}.
\label{FLRW_coordinate_distance_time_relation_for_redshift_derivation}
\end{equation}
Thus, we gained a relation between the coordinate distance and the time,
setting $r_{0}=0$ we obtain 
\begin{equation}
\int_{0}^{r_{1}}\frac{dr}{\sqrt{1-kr^{2}}}=f(r_{1})=\int_{t_{1}}^{t_{0}} \frac{dt}{S(t)}.  \label{FLRW_redshift_function_of_distance_relation}
\end{equation}
Shifting the time of the emission $t_{1}\rightarrow t_{1}+\delta t_{1}$ will
result in a shift of the detection time, i.e.\ $t_{0}\rightarrow t_{0}+\delta
t_{0}$. Since the lhs of (\ref{FLRW_redshift_function_of_distance_relation})
does not depend on the emission or detection time we can infer 
\begin{eqnarray}
\int_{t_{1}}^{t_{0}}\frac{dt}{S(t)}=\int_{t_{1}+\delta t_{1}}^{t_{0}+\delta
t_{0}}\frac{dt}{S(t)}\, &\Leftrightarrow &\,\int_{t_{0}}^{t_{0}+\delta t_{0}} \frac{dt}{S(t)}=\int_{t_{1}}^{t_{1}+\delta t_{1}}\frac{dt}{S(t)}
\label{FLRW_redshift_integration_limits_shifted} \\
\delta t_{0,1}\ll t_{0,1}\rightarrow S(t)=\texttt{\small \rm const} &\Rightarrow &\frac{
\delta t_{0}}{S(t_{0})}=\frac{\delta t_{1}}{S(t_{1})}.
\label{FLRW_delta_t_scale_relation}
\end{eqnarray}
Interpreting $\delta t_{0,1}$ as the times between two wave crests at
emission and at detection, i.e.\ relating them to the wavelength $\delta
t_{0,1}\backsim \lambda _{0,1}$, one obtains 
\begin{equation}
\frac{\lambda _{1}}{\lambda _{0}}=\frac{S(t_{1})}{S(t_{0})}.
\label{FLRW_wavelengths_redshift_definition}
\end{equation}
Using the common astronomical definition of the redshift, i.e.\
\begin{equation}
z:=\frac{\lambda _{0}}{\lambda _{1}}-1\rightarrow \left\{\begin{small} 
\begin{tabular}{ll}
$>0$&\texttt{\small \rm  redshifted}\\
$<0$&\texttt{\small \rm  blueshifted}
\end{tabular}\end{small}
\right., 
\end{equation}
we find the following relation
between the scale factor and the redshift in a Robertson-Walker spacetime 
\begin{equation}
1+z=\frac{S(t_{0})}{S(t_{1})}.
\label{FLRW_redshift_scale_factor_relation_final}
\end{equation}
Thus, if the universe is expanding, then distant sources should be
redshifted. Note that the derivation presented above makes use of three
crucial assumptions: (i) the paths of photons are completely determined by the
metric, (ii) the dispersion relation $\lambda \nu =c$ is valid at all times,
and (iii) the observed wavelengths are small compared to the size of the universe.

Let us now derive an alternative form of the Friedman
equation. For that purpose we make use of the expression for the redshift
(\ref{FLRW_redshift_scale_factor_relation_final}), and the relation between
the scale factor and the density
(\ref{FLRW_energy_density_scale_factor_relation_for_arbitrary_w_const}). With
$\Omega _{k}:=-\frac{k}{S^{2}H^{2}}$ equation (\ref{FLRW_first_friedman_written_with_omega_total}) turns into
\begin{equation}
\Omega_w +\Omega _{\lambda }+\Omega _{k}=1.  \label{FLRW_Omega_sums_equal_1}
\end{equation}
For normal matter ($w=0$) we can rewrite the Friedman equation as follows 
\begin{eqnarray}
&H^{2}=&\frac{\kappa }{3}\mu _{\texttt{\tiny \rm \rm m}}-\frac{k}{S^{2}}+\frac{\lambda }{3}
\quad \stackrel{(\ref{FLRW_energy_density_scale_factor_relation_for_arbitrary_w_const})}{\Rightarrow}
\quad H^{2}=\frac{\kappa }{3S^{3}}-\frac{k}{S^{2}}+\frac{\lambda }{3}  \nonumber ,\\
\stackrel{(\ref{FLRW_redshift_scale_factor_relation_final})}{\Rightarrow}& H^{2}=&H_{0}^{2}\left[ \frac{\kappa }{3H_{0}^{2}}\mu _{\texttt{\tiny \rm \rm m}0}\,\left(1+z\right) ^{3}-\frac{k}{S_{0}^{2}H_{0}^{2}}\,\left( 1+z\right) ^{2}+\frac{\lambda }{3H_{0}^{2}}\right]   \nonumber ,\\
\stackrel{(\ref{FLRW_Omega_sums_equal_1})}{\Rightarrow}&H^{2}=&H_{0}^{2}\left[ \left( 1+z\right) ^{2}\left( 1+z\,\Omega _{\texttt{\tiny \rm \rm m}0}\right) \,-z\,\left( 2+z\right) \,\,\Omega _{\lambda 0}\,\right] .
\label{FLRW_Hubble_factor_in_terms_of_Omega_Lambda_Omega_m}
\end{eqnarray}
Here and throughout the rest of the paper we denote present values of
certain quantities, i.e. at the time $t_0$, by an index $0$. The Hubble parameter in terms of the redshift reads 
\begin{eqnarray}
&&
\hspace{-2cm}
H=\frac{d}{dt}\log \left( \frac{S}{S_{0}}\right) =\frac{d}{dt}\log \left( 
\frac{1}{1+z}\right) =-\frac{1}{1+z}\frac{dz}{dt}
\label{FLRW_relation_between_dz_dr_dt_and_H} \\
&
\stackrel{(\ref{FLRW_Hubble_factor_in_terms_of_Omega_Lambda_Omega_m})}{\Rightarrow}
&\frac{dt}{dz}=-H_{0}^{-1}\left( 1+z\right) ^{-1}\left[ \left( 1+z\right)
^{2}\left( 1+z\,\Omega _{\texttt{\tiny \rm \rm m}0}\right) \,-z\,\left( 2+z\right)\,\,\Omega _{\lambda 0}\right] ^{-\frac{1}{2}}.
\label{FLRW_dr_dt_as_combination_of_the_density_parameters}
\end{eqnarray}
Before we proceed with the derivation of the luminosity distance in the FLRW
model, we note that there is a remarkable connection between the density
parameters and the so-called deceleration parameter $q$. The deceleration
parameter, which is commonly introduced when one expands the scale factor
around a certain time (cf.\ \cite{KolbTurner}), is defined as follows 
\begin{equation}
q:=-\frac{\ddot{S}S}{\dot{S}^{2}}=-\frac{\ddot{S}}{H^{2}S}=\frac{d}{dt}
H^{-1}-1.
\label{OBSERVABLES_definition_of_the_deceleration_factor_in_terms_of_the_hubble_parameter}
\end{equation}
Thus, in case of a FLRW model which contains only usual matter and a
cosmological constant, cf.\ equation (\ref{FLRW_Hubble_factor_in_terms_of_Omega_Lambda_Omega_m}), the deceleration
factor at present time is given by the simple expression 
\begin{equation}
q_{0}=\frac{\Omega _{\texttt{\tiny \rm \rm m}0}}{2}-\Omega _{\lambda 0}.
\label{OBSERVABLES_present_day_deceleration_factor_for_lambda_m_model}
\end{equation}

\subsection{Luminosity distance\label{OBSERVABLES_Luminosity_distance_section}}

In order to assign a distance to objects one introduces the so-called
luminosity distance, using the fact that the light from objects far away
from us appears fainter than the light from nearby ones. The astronomical
definition reads 
\begin{eqnarray}
d_{\texttt{\tiny \rm \rm luminosity}} &:=&\left( \frac{\texttt{\small \rm
      energy per time produced by source}}{\texttt{\small \rm energy per time per area detected by observer}}\right) ^{\frac{1}{2}}  \nonumber \\
&=&\left( \frac{\texttt{\small \rm luminosity}}{4\pi \times \texttt{\small \rm flux}}\right) ^{\frac{1}{2}}=\left( \frac{\breve{L}}{4\pi \breve{F}}\right) ^{\frac{1}{2}}.
\label{OBSERVABLES_definition_luminosity_distance}
\end{eqnarray}
Thus, by measuring $\breve{F}$ and with knowledge of $\breve{L}$ (via a
standard candle, a supernova e.g.) we are able to determine the distance
$d_{\texttt{\tiny \rm luminosity}}$. Of course this distance definition
implies that we know how much light is emitted by the source at least during a
specific time interval. We will not investigate this question here any
further since the search for an appropriate model of the source belongs to the realm of
astrophysics (a discussion of type Ia supernova models can be found in \cite{Niemeyer}). The question we have to ask ourselves is:
How is the luminosity distance related to parameters within the FLRW model?
Since energy is conserved the following equation is supposed to hold 
\begin{equation}
\breve{L}\,\,\delta t_{1}\,\delta \lambda _{1}=\breve{F}\,\,\delta
t_{0}\delta \lambda _{0}A_{0},  \label{FLRW_initial_observed_energy}
\end{equation}
here $A$ denotes the area of the 2-sphere at the detection time $t=t_{0}$,
and $\delta t_{0,1}$, and $\delta \lambda _{0,1}$ the different length and
time scales at emission and detection due to global expansion. The distance $d_{\texttt{\tiny \rm \rm FLRW}}$ between an object at $r=r_{1}$, which emits light at $t=t_{1},$ and an observer at $r=r_{0}=0$, who detects the light at $t=t_{0}$, is given by 
\begin{equation}
d_{\texttt{\tiny \rm \rm FLRW}}=S(t_{0})r_{1}\quad \Rightarrow \quad A_{0}=4\pi d_{\texttt{\tiny \rm \rm FLRW}}^{2}=4\pi S^{2}(t_{0})r_{1}^{2}.
\label{FLRW_comoving_distance_surface_of_the_2_sphere}
\end{equation}
Thus, with the help of equations (\ref{FLRW_wavelengths_redshift_definition})--(\ref{FLRW_redshift_scale_factor_relation_final}) the observed flux can
be expressed in the following form 
\begin{equation}
\breve{F} =\frac{\breve{L}}{4\pi S^{2}(t_{0})r_{1}^{2}}\frac{\delta
  t_{1}}{\delta t_{0}}\frac{\delta \lambda _{1}}{\delta \lambda _{0}} \stackrel{(\ref{FLRW_wavelengths_redshift_definition})+(\ref{FLRW_redshift_scale_factor_relation_final})}{=} \frac{\breve{L}}{4\pi S^{2}(t_{0})r_{1}^{2}}(1+z)^{-2}.
\label{FLRW_luminosity_distance_final}
\end{equation}
Comparison with (\ref{OBSERVABLES_definition_luminosity_distance}) yields 
\begin{equation}
d_{\texttt{\tiny \rm \rm luminosity}}=S(t_{0})\,r_{1}\,\left( 1+z\right) .
\label{FLRW_distance_redshift_relation_final_with_r_1}
\end{equation}
Additionally one wants to replace $r_{1}$ by the scale factor $S$. We make
use of equation (\ref{FLRW_redshift_function_of_distance_relation}) and obtain 
\begin{eqnarray}
&f(r_{1}) =\int_{0}^{r_{1}}\frac{dr}{\sqrt{1-kr^{2}}}=\int_{t_{1}}^{t_{0}} \frac{dt}{S(t)}=\left\{ 
\begin{tabular}{lll}
$\arcsin \left( r_{1}\right) $ &  & $k=+1$ \\ 
$r_{1}$ & for & $k=0$ \\ 
$\arcsin $h$\left( r_{1}\right) $ &  & $k=-1$
\end{tabular}
\right. , \nonumber \\
\Rightarrow &r_{1}=\left\{\begin{tabular}{lll}
$\sin \left( \int_{t_{1}}^{t_{0}}\frac{dt}{S(t)}\right) $ &  & $k=+1$ \\ 
$\int_{t_{1}}^{t_{0}}\frac{dt}{S(t)}$ & for & $k=0$ \\ 
$\sin $h$\left( \int_{t_{1}}^{t_{0}}\frac{dt}{S(t)}\right) $ &  & $k=-1$
\end{tabular}
\right. , \nonumber \\
\Rightarrow &d_{\texttt{\tiny \rm \rm luminosity}}=S(t_{0})\,\,\left( 1+z\right) \times
\left\{ 
\begin{tabular}{lll}
$\sin \left( \int_{t_{1}}^{t_{0}}\frac{dt}{S(t)}\right) $ &  & $k=+1$ \\ 
$\int_{t_{1}}^{t_{0}}\frac{dt}{S(t)}$ & for & $k=0$ \\ 
$\sin $h$\left( \int_{t_{1}}^{t_{0}}\frac{dt}{S(t)}\right) $ &  & $k=-1$
\end{tabular}
\right. .
\label{FLRW_luminosity_distance_r1_as_a_function_of_the_scale_factor}
\end{eqnarray}
Thus, we are able to express $r_{1}$ via the scale factor. In order to
derive $r_{1}$ explicitly we need a solution of the Friedman equations,
which of course depends on the cosmological model, i.e.\ the choice of the
parameters in (\ref{friedman1})--(\ref{friedman2}), we decide to
consider. Note that we did not make use of the field equations of the
underlying gravity theory up to this point. This fact will be crucial when we
derive an expression for the luminosity distance within a cosmological model which is not based on general relativity in the next section. As we will show
in the following subsection there is an elegant way to rewrite the luminosity
 distance in terms of the density parameters, which makes use of the expression for the
Hubble parameter as derived in equation
(\ref{FLRW_Hubble_factor_in_terms_of_Omega_Lambda_Omega_m}). Note that this is
the point when the field equations come into play.    

\paragraph{Special case\label{FLRW_special_case_subsection_for_luminosity_distance_in_case_of_omega_Lambda_omega_matter}}

In case of a Friedman model which contains only normal matter and a
contribution from the cosmological constant one can express the luminosity
distance as a function redshift and the model parameters. As we will show below, this expression is used when one wants to perform fits to observational data. Again we make use of the Robertson-Walker line element, i.e.\ 
\begin{eqnarray}
&&\frac{dr}{\sqrt{1-kr^{2}}}=\frac{dt}{S}
\quad \stackrel{(\ref{FLRW_redshift_scale_factor_relation_final})}{\Leftrightarrow}
\quad 
\frac{S_{0}}{\sqrt{1-kr^{2}}}dr=\left( 1+z\right) \,dt , \nonumber \\
&
\stackrel{(\ref{FLRW_dr_dt_as_combination_of_the_density_parameters})}{\Rightarrow}
&S_{0}\int_{0}^{r_{1}}\frac{dr}{\sqrt{1-kr^{2}}}=H_{0}^{-1}\int_{0}^{z_{1}}\frac{dz}{\sqrt{\left( 1+z\right) ^{2}\left( 1+z\,\Omega _{\texttt{\tiny \rm \rm m}0}\right)
-z\left( 2+z\right) \,\Omega _{\lambda 0}}},  \nonumber \\
&\Rightarrow &\Theta ^{-1}[r_{1}]=\left( H_{0}S_{0}\right)
^{-1}\int_{0}^{z_{1}}\frac{dz}{\sqrt{\left( 1+z\right) ^{2}\left(
1+z\,\Omega _{\texttt{\tiny \rm \rm m}0}\right) -z\left( 2+z\right) \,\Omega _{\lambda 0}}} 
, \nonumber \\
&\Rightarrow &d_{\texttt{\tiny \rm \rm luminosity}}=S_{0}\,\,\left( 1+z\right) \,\Theta \left[ \left( H_{0}S_{0}\right) ^{-1}\int_{0}^{z}F\left[ \tilde{z}\right] d \tilde{z}\right] .
\label{FLRW_luminosity_distamce_as_a_function_of_S_0_H_0_omega_lambda_omega_k}
\end{eqnarray}
Where we made use of the following definitions\label{OBSERVABLES_theta_page_marker} $\Theta \lbrack x]:=\left\{ 
\begin{tabular}{lll}
$\sin \left( x\right) $ &  & $k=+1$ \\ 
$x$ & for & $k=0$ \\ 
$\sin $h$\left( x\right) $ &  & $k=-1$
\end{tabular}
\right. $ and $F\left[ \tilde{z}\right] :=\left[ \left( 1+\tilde{z}\right)
^{2}\left( 1+\tilde{z}\,\Omega _{\texttt{\tiny \rm \rm m}0}\right)
-\tilde{z}\left( 2+\tilde{z}\right) \,\Omega _{\lambda 0}\right]
^{-\frac{1}{2}}$. If we make use of the definition of $\Omega _{k}$ we can rewrite equation (\ref
{FLRW_luminosity_distamce_as_a_function_of_S_0_H_0_omega_lambda_omega_k}) as
follows: 
\begin{eqnarray}
d_{\texttt{\tiny \rm \rm luminosity}} &=&\frac{\left( 1+z\right) }{H_{0}\sqrt{\left| \Omega
_{k0}\right| }}\,\Theta \left[ \sqrt{\left| \Omega _{k0}\right| }
\int_{0}^{z}F\left[ \tilde{z}\right] d\tilde{z}\right]  \nonumber \\
&
\stackrel{(\ref{FLRW_Omega_sums_equal_1})}{=}
&\frac{\left( 1+z\right) }{H_{0}\sqrt{\left| 1-\Omega _{\texttt{\tiny \rm \rm m}0}-\Omega
_{\lambda 0}\right| }}\,\Theta \left[ \sqrt{\left| 1-\Omega _{\texttt{\tiny \rm \rm m}0}-\Omega _{\lambda 0}\right| }\int_{0}^{z}F\left[ \tilde{z}\right] d\tilde{z}\right] .
\label{FLRW_luminosity_distance_final_as_a_function_of_H_0_Omega_lam_Omega_m}
\end{eqnarray}
Thus, within a Friedman model with normal matter and cosmological constant
the luminosity distance turns out to be function of the corresponding
density parameters, the Hubble constant, and the redshift, i.e.\ $d_{\texttt{\tiny \rm \rm luminosity}}=d_{\texttt{\tiny \rm \rm luminosity}}\left( z,H_{0},\Omega _{\texttt{\tiny \rm \rm m}0},\Omega
_{\lambda 0}\right) $. Note that this is a remarkable result since
$d_{\texttt{\tiny \rm \rm luminosity}}$ depends only on the present day values
  of the parameters within the model and the redshift.

\subsection{Magnitude-redshift relation \label{FLRW_magnitude_redshift_relation_section}}

Due to historical reasons astrophysicists often use the so-called magnitude as
unit for the luminosity of a stellar object. The relation between the
distance-redshift relation and the so-called magnitude-redshift relation (cf.\ 
\cite{PadmanabhanAstro2,Perlmutter,Perlmutter2}) is given by 
\begin{eqnarray}
\hspace{-1cm}m(z,H_{0},\Omega _{0},\Omega _{\lambda 0},w,M)&:=&M+5\log \left( \frac{d_{\texttt{\tiny \rm luminoity}}}{\texttt{\small \rm length}}\right) +25  \nonumber \\
&=&M+5\log \left( H_{0}\,d_{\texttt{\tiny \rm \rm luminosity}}\right) - 5\log \left( \frac{H_{0}}{\texttt{\small \rm length}}\right) +25.  \label{FLRW_magnitude_redshift_relation}
\end{eqnarray}
Where $M$ represents the absolute magnitude of the observed star. By
introducing a new constant $\mathcal{M}:=M-5\log H_{0}+25$ we are able to express the
distance-redshift relation in equation (\ref{FLRW_luminosity_distance_final_as_a_function_of_H_0_Omega_lam_Omega_m}) in a compact way as magnitude-redshift relation 
\begin{eqnarray}
\hspace{-2cm}m(z,H_{0},\Omega _{\texttt{\tiny \rm \rm m}0},\Omega _{\lambda
  0},M)\nonumber\\
\hspace{-0.5cm}=\mathcal{M}+5\log\left( \,\frac{\left( 1+z\right) }{\sqrt{\left| 1-\Omega _{\texttt{\tiny \rm \rm m}0}-\Omega_{\lambda 0}\right| }}\,\Theta \left[ \sqrt{\left| 1-\Omega _{\texttt{\tiny \rm \rm m}0}-\Omega _{\lambda 0}\right| }\int_{0}^{z}F\left[ \tilde{z}\right] d\tilde{z}\right] \right) .
\label{FLRW_exact_expression_for_magnitude_redshift_in_case_of_a_model_with_matter_cosmological_constant}
\end{eqnarray}
This relation is commonly used to extract cosmological parameters, like the
density parameters associated with normal matter and the cosmological
constant, by performing fits to data sets which were produced by the 
observation of standard candles, i.e.\ objects of known absolute magnitude. Equation (\ref{FLRW_exact_expression_for_magnitude_redshift_in_case_of_a_model_with_matter_cosmological_constant})
will be of use in section \ref{Numerical_results_section} where we will
perform fits to a real data set. Note that table \ref{tabelle_1} contains a
collection of all assumptions made during the derivation of the
magnitude-redshift relation within the cosmological standard model. 

\begin{table}
\caption{Assumptions made up to this point.}
\label{tabelle_1}
\begin{indented}
\item[]\begin{tabular}{@{}ll}
\br
Ansatz/Assumption & Equation \\ 
\mr
General Relativity as underlying gravity theory&(\ref{field_eq_fluid_classic})\\
Metric is of Robertson-Walker type (i.e.\ homogeneity and isotropy)& (\ref{spherical_coframe})--(\ref
{FLRW_robertson_walker_line_element}) \\ 
Photons follow null curves (i.e.\ are determined by the metric) & (\ref
{FLRW_coordinate_distance_time_relation_for_redshift_derivation}) \\ 
Observed wavelengths small compared to size of the universe & (\ref
{FLRW_delta_t_scale_relation}) \\ 
Dispersion relation valid at all times, and peculiar movement &\\
of the source neglect able & (\ref
{FLRW_wavelengths_redshift_definition}) \\ 
Sources of known constant absolute magnitude&(\ref{FLRW_magnitude_redshift_relation})\\  
Photons travel unimpeded between source and observer, i.e.\ no&\\
 gravitational potentials or dust between source and observer&(\ref{spherical_coframe})--(\ref{classical_ideal_fluid_cosmology})\\
\br
\end{tabular}
\end{indented}
\end{table}

\section{Magnitude-redshift relation within the Weyl-Cartan model\label{EXTWEYL_magnitude_redshift_relation}}

After our review of the derivation of the magnitude-redshift relation within
the cosmological standard model we will now switch to the new Weyl-Cartan
model which was presented in \cite{Extweyl}. We will not discuss the details of this model at this point, therefore the reader should
consult sections 2--4 of \cite{Extweyl} in order to get an idea which
assumptions were made during the derivation of the field equations of this
model.

\subsection{Field equations\label{Field_eq_in_ext_model_section}}

Let us collect the field equations derived in \cite{Extweyl} (cf.\ eqs.\
(56)-(62)). Note that we make use of the form in which the constant $\Xi$,
which was introduced in eq.\ (54) of \cite{Extweyl}, is already set
to zero. 

\begin{eqnarray}
\chi \left( \left( \frac{\dot{S}}{S}\right) ^{2}+\frac{k}{S^{2}}\right)
-( a_{4} &+ a_{6} ) \kappa \left( \left( \frac{\ddot{S}}{S}\right)
^{2}-\left( \left( \frac{\dot{S}}{S}\right) ^{2}+\frac{k}{S^{2}}\right)
^{2}\right)  \nonumber \\
&=\kappa \left( p_{r}-4c\left( \frac{\zeta }{S}\right) ^{2}\right) ,
\label{field_final_zeta_XI_<>_0_1} \\
\chi \left( \Lambda +\frac{\ddot{S}}{S}\right) +\left( a_{4}+a_{6}\right)
&\kappa \left( \left( \frac{\ddot{S}}{S}\right) ^{2}-\left( \left( \frac{\dot{S}}{S}\right) ^{2}+\frac{k}{S^{2}}\right) ^{2}\right)  \nonumber \\
&=-\kappa \left( p_{r}-4c\left( \frac{\zeta }{S}\right) ^{2}\right) ,
\label{field_final_zeta_XI_<>_0_2} \\
\frac{\ddot{S}}{S}+\left( \frac{\dot{S}}{S}\right) ^{2}+\frac{k}{S^{2}}
&=\Lambda ,  \label{field_final_zeta_XI_<>_0_3} \\
4\frac{\dot{S}}{S}\mu +\dot{\mu}\ &=8c\left( \frac{\zeta }{S}\right)
^{2}\left( \frac{\dot{S}}{S}+\frac{\dot{\zeta}}{\zeta }\right) .
\label{field_final_zeta_XI_<>_0_4}
\end{eqnarray}
Additionally, we have the following relation between the pressure and the
energy-density 
\begin{equation}
\mu =3p_{r}-8c\left( \frac{\zeta }{S}\right) ^{2}.
\label{xi_=_0_energy_and_stresses_relation}
\end{equation}
As one can see from equations
(\ref{field_final_zeta_XI_<>_0_1})--(\ref{xi_=_0_energy_and_stresses_relation})
there is an additional function $\zeta(t)$ entering the field equations, besides of the scale factor $S(t)$ from the Robertson-Walker
metric. This function stems from an ansatz for the Weyl 1-form $Q$ which
governs the non-Riemannian features of the model, cf.\ section 4 of
\cite{Extweyl}. Here $k$ denotes the usual parameter within the
Robertson-Walker metric, $\chi, a_4, a_6, c,$ and $\kappa$ are coupling
constants, and $\Lambda$ is the so-called induced cosmological constant.    

\subsection{Magnitude-redshift relation\label{Magnitude_redshift_section}}

As we have shown in sections
\ref{OBSERVABLES_Luminosity_distance_section} and \ref{FLRW_magnitude_redshift_relation_section},
the magnitude-redshift relation depends on the mechanism of light propagation of the underlying
theory of gravity and the field equations. Since we now work in a space which carries
nonmetricity, besides of the torsion which does not affect light propagation, the question arises if the trajectories of photons deviate from the
ones in the Riemannian theory. The answer to this question is no, since in a Weyl-Cartan spacetime the null geodesics are determined by the usual geodesic equation as in the Riemannian case. Hence we are allowed to use
relation (\ref{FLRW_relation_between_dz_dr_dt_and_H}), which is a consequence
of the Robertson-Walker metric, in order to derive the magnitude-redshift relation.
Thus, our next aim is to express the Hubble function in terms of some
density functions and model parameters. From equation (\ref
{field_final_zeta_XI_<>_0_3}) we infer 
\begin{equation}
H^{2}=\Lambda -\frac{\ddot{S}}{S}-\frac{k}{S^{2}}.
\label{EXTWEYL_third_field_to_express_hubble}
\end{equation}
In order to eliminate the second order term we make use of equation (\ref
{field_final_zeta_XI_<>_0_2}). Reinserting the solution for $\ddot{S}/S$
from this equation into (\ref{EXTWEYL_third_field_to_express_hubble}) leads
to 
\begin{eqnarray}
H^{2} &=&\frac{\left( \Lambda ^{2}S^{2}-2\Lambda k\right) \,\kappa \,\left(
a_{4}+a_{6}\right) -\chi k+\kappa S^{2}p_{r}-4\kappa c\zeta ^{2}}{\left(
2\kappa \Lambda \left( a_{4}+a_{6}\right) +\chi \right) S^{2}} \\
&=&\frac{H_{0}^{2}}{\left( 2\kappa \Lambda \left( a_{4}+a_{6}\right) +\chi
\right) }\left\{ \kappa \left( a_{4}+a_{6}\right) \left[ \frac{\Lambda ^{2}}{H_{0}^{2}}-\frac{2\Lambda k}{H_{0}^{2}S_{0}^{2}}\left( \frac{S_{0}}{S}\right) ^{2}\right] \right.  \nonumber \\
&&\left. -\frac{\chi k}{H_{0}^{2}S_{0}^{2}}\left( \frac{S_{0}}{S}\right)
^{2}+\frac{\kappa p_{r}}{H_{0}^{2}}-\frac{4\kappa c\zeta ^{2}}{
H_{0}^{2}S_{0}^{2}}\left( \frac{S_{0}}{S}\right) ^{2}\right\} \\
&
\stackrel{(\ref{FLRW_redshift_scale_factor_relation_final})}{=}
&\frac{H_{0}^{2}}{\left( 2\kappa \Lambda \left( a_{4}+a_{6}\right) +\chi
\right) }\left\{ \frac{}{}\kappa \left( a_{4}+a_{6}\right) \left[ H_{0}^{2}\Omega
_{\Lambda 0}^{2}-2\Lambda \Omega _{k0}\,\left( 1+z\right) ^{2}\right] \right.
\nonumber \\
&&\left. -\chi \Omega _{k0}\left( 1+z\right) ^{2}+\frac{\kappa p_{r}}{
H_{0}^{2}}-\frac{4\kappa c\zeta ^{2}}{H_{0}^{2}S_{0}^{2}}\left( 1+z\right)
^{2}\right\} .  \label{EXTWEYL_H_squared_general_form}
\end{eqnarray}
In the last equation we introduced the density parameters $\Omega
_{k}:=\frac{k}{H^{2}S^{2}},$ and $\Omega _{\Lambda }:=\frac{\Lambda }{H^{2}}$. Subsequently, we
have to choose an equation of state and an appropriate ansatz for $\zeta $.
We choose the eos to be of the form $p = w \mu$ with $w={\rm const}$, and make use of the solution
for $\mu $ obtained in \cite{Extweyl} equation (65), i.e.\ $\mu =-\frac{8c}{1-3w}\left( \frac{\zeta }{S}\right) ^{2}$. Hence we can infer 
\begin{eqnarray}
H^{2} &=&\frac{H_{0}^{2}}{\left( 2\kappa \Lambda \left( a_{4}+a_{6}\right)
+\chi \right) } \left\{ \frac{}{}\kappa \left( a_{4}+a_{6}\right) \left[
H_{0}^{2}\Omega _{\Lambda 0}^{2}-2\Lambda \Omega _{k0}\,\left( 1+z\right)
^{2}\right] \right.  \nonumber \\
&&\left. -\chi \Omega _{k0}\left( 1+z\right) ^{2}-\frac{4\kappa c\zeta ^{2}}{
H_{0}^{2}S_{0}^{2}}\left( \frac{1-w}{1-3w}\right) \left( 1+z\right) ^{2}\right\} .
\label{EXTWEYL_H_squared_with_inserted_equation_of_state_and_density}
\end{eqnarray}
For $\zeta $ we make use of the solution mentioned in \cite{Extweyl}
equation (70), i.e.\ $\zeta =\iota /S$ with $\iota =$ const, which finally
yields 
\begin{eqnarray}
\hspace{-1cm}H^{2} 
=\frac{H_{0}^{2}}{\left( 2\kappa H_{0}^{2}\Omega _{\Lambda 0}\left(
a_{4}+a_{6}\right) +\chi \right) }\left\{ \frac{}{}\kappa \left( a_{4}+a_{6}\right)
\left[ H_{0}^{2}\Omega _{\Lambda 0}^{2}-2H_{0}^{2}\Omega _{\Lambda 0}\Omega
_{k0}\,\left( 1+z\right) ^{2}\right] \right.  \nonumber \\
\left. -\chi \Omega _{k0}\left( 1+z\right) ^{2}-4\bigskip \Omega _{\zeta
0}\left( 1+z\right) ^{4}\left( \frac{1-w}{1-3w}\right) \right\} ,
\label{EXTWEYL_H_squared_final_with_all_density_parameters}
\end{eqnarray}
where we introduced the new density parameter \bigskip $\Omega
_{\zeta }:=\frac{\kappa c\iota ^{2}}{H^{2}S^{4}}.$

\paragraph{Special case \label{EXTWEYL_Magniute_redshift_for_lambda_=_0}}

In case of a vanishing induced cosmological constant, i.e.\ $\Lambda =0$,
equation (\ref{EXTWEYL_H_squared_final_with_all_density_parameters}) turns
into 
\begin{equation}
H^{2}=\frac{H_{0}^{2}}{\chi }\left[ 4\left( 1+z\right) ^{4}\Omega _{\zeta
0}\left( \frac{w-1}{1-3w}\right) -\chi \Omega _{k0}\left( 1+z\right) ^{2}
\right] .  \label{EXTWEYL_H_squared_for_mag_redshift_for_lambda_0_}
\end{equation}
In order to calculate the magnitude-redshift relation we remember equation (\ref{FLRW_relation_between_dz_dr_dt_and_H}) and insert this expression into
the first line of equation (\ref
{FLRW_luminosity_distamce_as_a_function_of_S_0_H_0_omega_lambda_omega_k}),
yielding the following luminosity distance 
\begin{eqnarray}
d_{\texttt{\tiny \rm \rm luminosity}}(z,H_{0},\Omega _{k0},\Omega _{\zeta 0},\chi ,w) =\frac{\left( 1+z\right) }{H_{0}\sqrt{\left| \Omega _{k0}\right|
    }}\Theta \left[ \sqrt{\frac{\left| \Omega _{k0}\right| }{\chi }}\int_{0}^{z}G[\tilde{z}]
d\tilde{z}\right] ,  \label{luminosity_distance_final_for_lambda_=_0_case}
\end{eqnarray}
with $\Theta $ as defined after eq.\
(\ref{FLRW_luminosity_distamce_as_a_function_of_S_0_H_0_omega_lambda_omega_k}),
and $G[\tilde{z}]:=\left[ 4\left( 1+\tilde{z}\right) ^{4}\Omega _{\zeta 0}\left( \frac{w-1}{1-3w}\right) -\chi \Omega _{k0}\left( 1+\tilde{z}\right) ^{2}\right] ^{-\frac{1}{2}}$. Hence the magnitude-redshift relation
is now given by (cf.\ eq.\ (\ref{FLRW_magnitude_redshift_relation})) 
\begin{eqnarray}
\hspace{-2cm}m(z,H_{0},\Omega _{\zeta 0},\Omega _{k0},w,M,\chi )&=\mathcal{M}
+5\log \left\{ \frac{}{} H_{0}\,d_{\texttt{\tiny \rm \rm luminosity}}(z,H_{0,}\Omega_{\zeta 0},\Omega _{k0},\chi ,w)
\right\} \nonumber \\
&=\mathcal{M}+5\log \left\{ \frac{\left( 1+z\right) }{\sqrt{\left| \Omega
_{k0}\right| }}\,\Theta \left[ \sqrt{\frac{\left| \Omega _{k0}\right| }{\chi 
}}\int_{0}^{z}G[\tilde{z}]\,d\tilde{z}\right] \right\} .
\label{EXTWEYL_magnitude_redshift_final_for_lambda_=_0_solutions}
\end{eqnarray}
Note that in contrast to the FLRW model there is no simple relation as in (\ref{FLRW_Omega_sums_equal_1})
between the density parameters. Therefore we cannot eliminate the density
parameter $\Omega_k$ in
the equation for the magnitude. In case of a flat model equation (\ref
{EXTWEYL_magnitude_redshift_final_for_lambda_=_0_solutions}) reduces to 
\begin{eqnarray}
\hspace{-2cm}m(z,H_0,\Omega _{\zeta 0},w,M,\chi )\nonumber\\
=\mathcal{M}+5\log \left\{ \frac{(1+z)}{\chi }\,\left[ \int_{0}^{z}\left[ 4\left( 1+\tilde{z}\right)
^{4}\Omega _{\zeta 0}\left( \frac{w-1}{1-3w}\right) \right] ^{-\frac{1}{2}}d\tilde{z}\right] \right\} .  \label{EXTWEYL_mr_flat_lambda_0}
\end{eqnarray}
In the next section we will make use of equation (\ref{EXTWEYL_magnitude_redshift_final_for_lambda_=_0_solutions}) in order to
determine whether it is possible to describe the available type Ia
supernova data within our model. Finally, we mention that the Hubble parameter found in equation (\ref
{EXTWEYL_H_squared_for_mag_redshift_for_lambda_0_}) enables us to derive
the deceleration parameter, which was defined in equation
(\ref{OBSERVABLES_definition_of_the_deceleration_factor_in_terms_of_the_hubble_parameter}). Its present day value is given by 
\begin{equation}
q_{0}=\frac{4\Omega _{\zeta 0}\left( w-1\right) }{4\left( w-1\right) \Omega
_{\zeta 0}+\chi \left( 3w-1\right) \Omega _{k0}}.
\label{EXTWEYL_present_day_deceleration_factor_in_the_lambda_zero_model}
\end{equation}

\section{Numerical results\label{Numerical_results_section}}

In this section we will present the numerical results obtained by fitting the
magnitude-redshift relations in (\ref{FLRW_exact_expression_for_magnitude_redshift_in_case_of_a_model_with_matter_cosmological_constant}) and (\ref{EXTWEYL_mr_flat_lambda_0}) 
to a real data set. We start with a collection of the different available data 
sets of type Ia supernovae, which were also used by other teams to determine the
cosmological parameters.

\subsection{Data sets\label{DATA_sets_section}}

In table \ref{tabelle_2} we collected the number of supernovae and the references which actually contain the data.
\begin{table}
\caption{SNIa data sets.}
\label{tabelle_2}
\begin{indented}
\item[]\begin{tabular}{@{}lllllll}
\br
Symbol & Number of SN & Reference & Comments &  &  &  \\ \mr
I & $18$ & p. 571, \cite{Hamuy, Perlmutter} & Cal\'{a}n/Tololo survey &  &  &  \\ 
II & $42$ & p. 570, \cite{Perlmutter} & Supernova Cosmology Project &  &  & 
\\ 
III & $10$ & p. 1021, \cite{Riess2} & High-z Supernova Search Team &  &  & 
\\ 
IV & $10$ & p. 1020, \cite{Riess2} & Same as III but MLCS method &  &  &  \\ 
V & $1$ & \cite{Riess1} & Farthest SNIa observed to date &  &  &  \\ 
VI & $27$ & p. 1035, \cite{Riess2} & Low-redshift MLCS/template &  &  & \\\br
\end{tabular}
\end{indented}
\end{table}
Note that the data sets of the different groups is not directly comparable.
Perlmutter \etal \cite{Perlmutter} provide the effective magnitude $m_{\texttt{\tiny \rm \rm B}}^{\texttt{\tiny \rm \rm eff}} $ in the B band, while Riess \etal \cite{Riess2} use
the so-called distance modulus $\mu $\footnote{Not to be confused with the
  energy-density within the field equations.}. As shown by Wang in \cite{Wang}
it is possible to find a relation between this two data sets by comparing the
data of 18 SNe Ia published by both groups. The definition of the magnitude as
given in equation (\ref{FLRW_magnitude_redshift_relation}) is compatible with
the definition used by Perlmutter \etal, it is related to the definition of Riess \etal by 
\begin{equation}
m=M+\mu =M+5\log d_{\texttt{\tiny \rm \rm luminosity}}+25=\mathcal{M}+5\log
H_{0}d_{\texttt{\tiny \rm \rm luminosity}}.  \label{definition_of_the_distant_modulus}
\end{equation}
As shown in \cite{Wang} we have to choose $M=-19.33\pm 0.25$ in order to
transform the different data sets into each other. Note that this value
corresponds to the MLCS method of Riess \etal. In the following we will
make use of the data of Wang which contains 92 data points and can be viewed as
a compilation of the sets I, II, IV, and VI from table \ref{tabelle_2} in
which some outliers were removed.   

\subsection{Fitting method}\label{NUMERICAL_fitting_method_section}

Since we want results which are comparable to the analysis of the
combined data set by Wang in \cite{Wang}, we are going to
minimize\footnote{Not to be confused with the coupling constant within the
  field equations in eqs.\ (\ref{field_final_zeta_XI_<>_0_1})--(\ref{field_final_zeta_XI_<>_0_4}).} 
\begin{equation}
\chi^2:=\sum\limits_{i=1}^{92}\frac{\left[\mu_i^{\rm theory}\left(z_i | {\tiny
        \rm parameters} \right)-\mu_i^{\rm measured}\right]^2}{\sigma^2_{\mu
        \,\, i}+\sigma^2_{mz \,\, i}}, \label{chi_square_fit_function}
\end{equation}
in order to obtain the best-fit parameters within the standard and the
Weyl-Cartan model \cite{Recipes,Bevington,Martin}. Here $\mu_i^{\rm theory}$ denotes the distance modulus at a
certain redshift $z_{i}$ as defined in
(\ref{definition_of_the_distant_modulus}). The error of the measured
$\mu_i^{\rm measured}$ is given by $\sigma^2_{\mu \,\, i}$. The dispersion
in the distance modulus $\sigma_{mz}$ due to the dispersion in the galaxy redshift,
$\sigma_z$, can be calculated iteratively by
\begin{equation}
\sigma_{mz}:=\frac{5}{ln10}\left[\frac{1}{d_{\rm luminosity}}\frac{\partial
    d_{\rm luminosity}}{\partial z}\right] \sigma_z \label{Sigma_z_as_in_Wang}
\end{equation}
according to Wang (cf.\ equation (13) of \cite{Wang}). We perform a brute
force calculation on grids as denoted in table \ref{tabelle_3} in order to
find the minimum of (\ref{chi_square_fit_function}).

\begin{table}
\caption{Grids used for minimization.}
\label{tabelle_3}
\begin{indented}
\item[]\begin{tabular}{@{}ll}
\br
 Parameters & [Interval, Stepsize] \\ \mr
 $\{\Omega _{\texttt{\tiny \rm \rm m}0},\Omega _{\lambda 0}\}$ & $\left\{
[-2\dots 4,0.01] , [-2\dots 4,0.01]\right\} $ \\ 
 $\{\Omega _{\texttt{\tiny \rm \rm k}0},\Omega _{\zeta 0},\chi ,w\}$ &
 $\left\{[-2\dots 4,0.01] , [-1\dots 1,0.01],1,[-1\dots 0,0.1]\right\}$ \\ 
\br 
\end{tabular}
\end{indented}
\end{table}

\subsection{Best-fit parameters\label{NUMERICAL_Best_fit_parameters_section}}

\begin{figure}[ht]
\setlength{\unitlength}{1mm}
\begin{picture}(80,50)
\epsfig{file=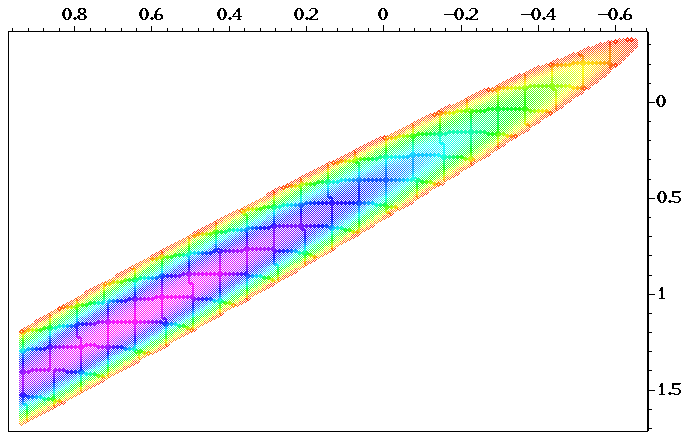}
\put(-78,26){\tiny{$\Omega_{\lambda0}$}}
\put(-40,2){\tiny{$\Omega_{m0}$}}
\put(-28,23){\tiny{$H_0=65$}}
\put(-28,20){\tiny{$\chi^2_{\rm min}=134.54$}}
\put(-28,35){\tiny{$68.3\%$}}
\put(-18,43){\tiny{$95.4\%$}}
\end{picture}
\begin{picture}(80,50)
\epsfig{file=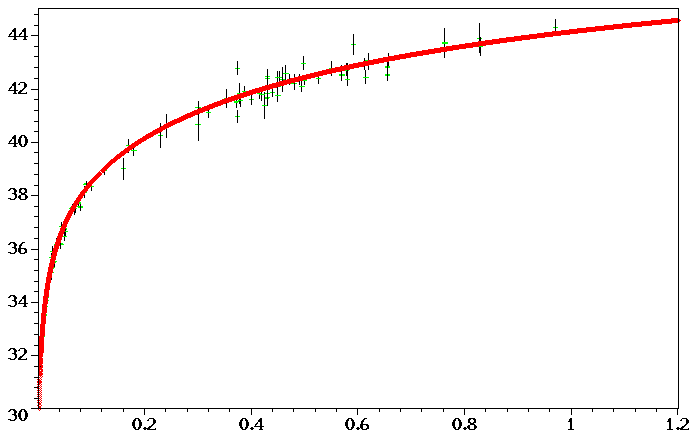}
\put(-78,26){\tiny{$\mu$}}
\put(-40,1){\tiny{$z$}}
\put(-22,38){\tiny{F13}}
\end{picture}
\begin{picture}(80,50)
\epsfig{file=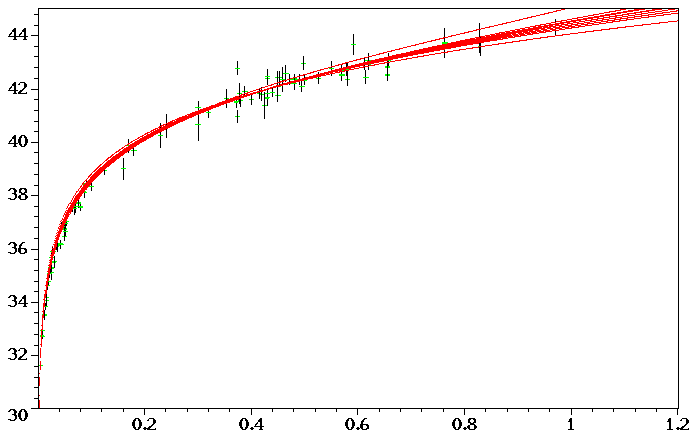}
\put(-78,26){\tiny{$\mu$}}
\put(-40,1){\tiny{$z$}}
\put(-40,30){\tiny{F1, $\dots$, F10, F13}}
\put(-40,20){\tiny{$k<0$}}
\end{picture}
\begin{picture}(80,50)
\epsfig{file=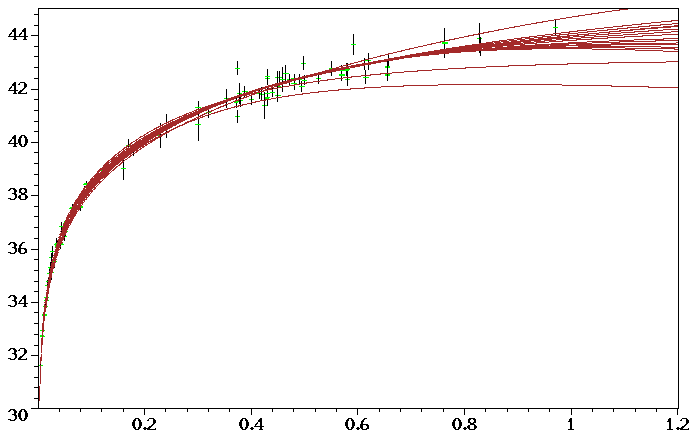}
\put(-78,26){\tiny{$\mu$}}
\put(-40,1){\tiny{$z$}}
\put(-40,30){\tiny{F11, F12, F14, $\dots$, F27}}
\put(-40,20){\tiny{$k>0$}}
\end{picture}
\caption[Standard chi distribution]{$\chi^2$-distribution in the
  $\left(\Omega_{{\rm m}0},\Omega_{\lambda 0}\right)$-plane and
  magnitude-redshift relation in case of the standard model. On the top lhs
  the $\chi^2$-distribution in the $H_0=65$ plane is shown, which also contains the
  best-fit parameter set F13. On the top rhs we plotted the distance modulus
  versus the redshift in case of the best-fit parameter set. The green dots
  correspond to the experimental data for 92 type Ia SNe as contained in
  the data set of Wang. In the two other figures we plotted the distance
  modulus versus the redshift for the other parameter sets in table \ref{tabelle_4}.}
\label{figure_1}
\end{figure}

\begin{figure}[ht]
\setlength{\unitlength}{1mm}
\begin{picture}(80,50)
\epsfig{file=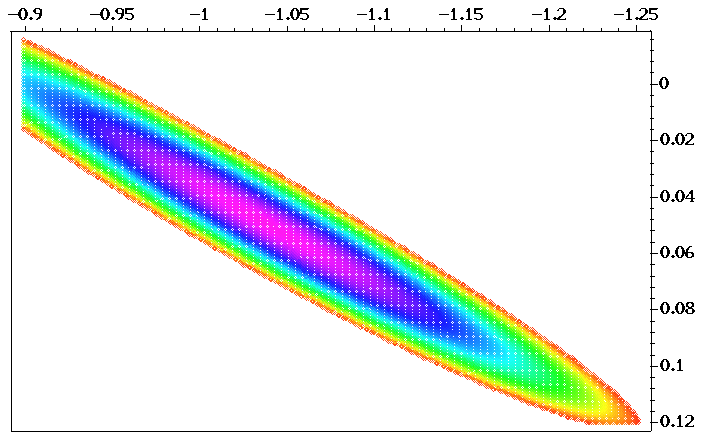}
\put(-78,24){\tiny{$\Omega_{\zeta0}$}}
\put(-40,1){\tiny{$\Omega_{k0}$}}
\put(-30,38){\tiny{$H_0=66$}}
\put(-30,35){\tiny{$\chi^2_{\rm min}=138.03$}}
\put(-60,40){\tiny{$95.4\%$}}
\put(-22,11){\tiny{$68.3\%$}}
\end{picture}
\begin{picture}(80,50)
\epsfig{file=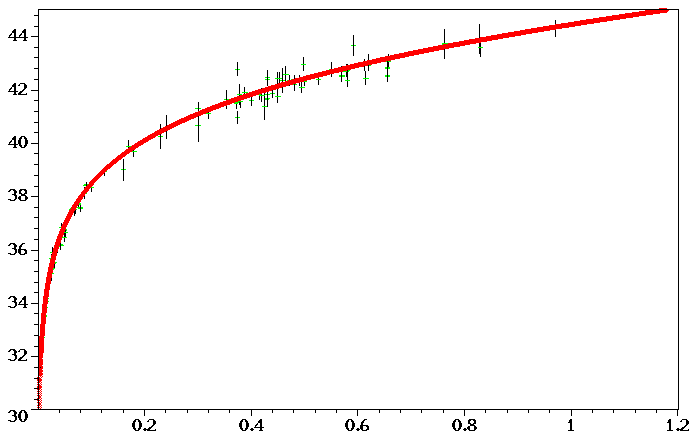}
\put(-78,26){\tiny{$\mu$}}
\put(-40,1){\tiny{$z$}}
\put(-40,30){\tiny{C1, $\dots$, C8}}
\put(-40,20){\tiny{$k<0$}}
\end{picture}
\caption[]{$\chi^2$-distribution in the $\left(\Omega_{k0},\Omega_{\zeta 0}\right)$-plane and magnitude-redshift relation
        in case of the Weyl-Cartan model. On the lhs the $\chi^2$-distribution in the $H_0=66$ plane is shown, which contains the
  best-fit parameter set C2. On the rhs we plotted the distance modulus
  versus the redshift in case of all parameter sets in table
        \ref{tabelle_5}. Since all of this sets fit the data very well the
        eight different curves appear as a single line at the selected resolution.}
\label{figure_2}
\end{figure}

In table \ref{tabelle_4} and \ref{tabelle_5} we collected the best-fit parameters
obtained by the method described in the previous section. Note that we did not
impose any constraints on our parameters, like spatial flatness, e.g., when performing our search.
In figure \ref{figure_1} and \ref{figure_2} plotted the corresponding distance
modulus versus redshift relation together with the data set of Wang which
consists out of 92 type Ia supernovae. Note that the $\chi^2$-distributions
displayed in figure \ref{figure_1} and \ref{figure_2} correspond to the
plane containing the best-fit parameters found on our grid. In both figures the $95.4\%$ confidence
level corresponds to the outer boundary. In table \ref{tabelle_6} and
\ref{tabelle_7} we collected some of the results of other groups which
performed a similar analysis within the FLRW as well as in non-standard
cosmological models. 
Note that these collections are by no means
exhaustive, we therefore apologize for not having mentioned all the works
which are devoted to this subject.
\begin{table}
\caption{Best-fit parameters (FLRW model).}
\label{tabelle_4}
\begin{indented}
\item[]\begin{tabular}{@{}lllllll}
\br
Symbol&$H_0$&      $\Omega _{\texttt{\tiny \rm \rm m}0}$&$\Omega _{\lambda
  0}$&$\chi^2$&$\chi^2_\nu$&$q_0$\\
\mr
F1&50&           0.45&       -2.00&        401.04&4.50&2.22\\
F2&54&          -0.60&       -2.00&        256.81&2.88&1.70\\
F3&55&          -0.81&       -2.00&        230.86&2.59&1.59\\
F4&56&          -1.00&       -2.00&        208.73&2.34&1.50\\
F5&57&          -1.18&       -2.00&        190.31&2.13&1.41\\
F6&58&          -1.34&       -2.00&        175.50&1.97&1.33\\
F7&59&          -2.00&       -2.00&        188.04&2.11&1.00\\
F8&60&          -1.15&       -1.49&        155.83&1.75&0.91\\
F9&61&          -0.64&       -0.79&        148.80&1.67&0.47\\
F10&62&         -1.14&       -1.21&        145.58&1.63&0.64\\
F11&63&          1.02&        1.29&        142.53&1.60&-0.78\\
F12&64&          3.10&        3.35&        233.03&2.61&-1.80\\\mr
F13&65&          0.63&        1.10&        134.54&1.51&-0.78\\\mr
F14&66&          0.80&        1.40&        134.91&1.51&-1.00\\
F15&67&          0.92&        1.63&        137.02&1.53&-1.17\\
F16&68&          0.99&        1.80&        140.98&1.58&-1.30\\
F17&69&          1.04&        1.94&        146.84&1.64&-1.42\\
F18&70&          1.07&        2.05&        154.66&1.73&-1.51\\
F19&72&          1.06&        2.17&        176.32&1.98&-1.64\\
F20&73&          1.04&        2.20&        190.23&2.13&-1.68\\
F21&74&          1.02&        2.23&        206.18&2.31&-1.72\\
F22&75&          0.98&        2.23&        224.21&2.51&-1.74\\
F23&76&          0.95&        2.24&        244.30&2.74&-1.76\\
F24&77&          0.91&        2.23&        266.47&2.99&-1.77\\
F25&78&          0.87&        2.22&        290.71&3.26&-1.78\\
F26&79&          3.05&        3.68&        733.20&8.23&-2.15\\
F27&80&          0.13&        1.45&        374.68&4.20&-1.38\\
\br
\end{tabular}
\item[] $[H_0]={\rm km \, s}^{-1}{\rm Mpc}^{-1}$.
\end{indented}
\end{table}

\begin{table}
\caption{Best-fit parameters (Weyl-Cartan model).}
\label{tabelle_5}
\begin{indented}
\item[]\begin{tabular}{@{}lllllllll}
\br
Symbol&$H_0$&$\Omega_{k0}$&$\Omega_{\zeta 0}$&$\chi$&$w$&$\chi^2$&$\chi^2_\nu$&$q_0$\\
\mr
C1&65&          -1.07&       0.05&       1.00&          -0.80&        138.034&1.56&-0.109\\
\mr
C2&66&          -1.03&       0.05&       1.00&          -0.90&        138.028&1.56&-0.110\\
\mr
C3&67&          -1.01&       0.03&       1.00&          -0.10&        138.034&1.56&-0.111\\
C4&69&          -0.95&       0.04&       1.00&          -0.60&        138.056&1.56&-0.106\\
C5&70&          -0.92&       0.04&       1.00&          -0.70&        138.034&1.56&-0.105\\
C6&71&          -0.90&       0.03&       1.00&          -0.20&        138.031&1.56&-0.111\\
C7&74&          -0.83&       0.04&       1.00&          -0.80&        138.039&1.56&-0.113\\
C8&80&          -0.71&       0.03&       1.00&          -0.50&        138.035&1.56&-0.112\\
\br
\end{tabular}
\item[] $[H_0]={\rm km \, s}^{-1}{\rm Mpc}^{-1}$.
\end{indented}
\end{table}

\begin{table}
\caption{Best-fit parameters other groups (FLRW model).}
\label{tabelle_6}
\begin{indented}
\item[]\begin{tabular}{@{}lllll}
\br
Symbol & Ref. & Best-fit parameters & Comment &  \\ \mr
S1 & \cite{Schmidt} & $\{\Omega _{\texttt{\tiny \rm \rm m}0}=-0.2,$ $\Omega _{\lambda 0}=0\}$
&   &  \\ 
S2 & \cite{Schmidt} & $\{\Omega _{\texttt{\tiny \rm \rm m}0}=0.4,$ $\Omega _{\lambda
0}=0.6\} $   &  \\ 
P1 & \cite{Perlmutter} & $\{\Omega _{\texttt{\tiny \rm \rm m}0}=0.28,$ $\Omega _{\lambda
0}=0.72\}$   &  \\ 
R1 & \cite{Riess2} & $\{\Omega _{\texttt{\tiny \rm \rm m}0}=0.24,$ $\Omega _{\lambda
0}=0.72\}$ & MLCS &  \\ 
R2 & \cite{Riess2} & $\{\Omega _{\texttt{\tiny \rm \rm m}0}=0.2,$ $\Omega _{\lambda 0}=0.8\}$
& Template &  \\ 
G1 & \cite{Garnavich1} & $\{\Omega _{\texttt{\tiny \rm \rm m}0}=-0.1,$ $\Omega _{\lambda
0}=0\}$ &  &  \\ 
G2 & \cite{Garnavich1} & $\{\Omega _{\texttt{\tiny \rm \rm m}0}=0.4,$ $\Omega _{\lambda
0}=0.6\}$ & MLCS &  \\ 
G3 & \cite{Garnavich1} & $\{\Omega _{\texttt{\tiny \rm \rm m}0}=0.3,$ $\Omega _{\lambda
0}=0.7\}$ & Template &  \\ 
V1 & \cite{Vishwakarma3} & $\{\Omega _{\texttt{\tiny \rm \rm m}0}=0.28,$ $\Omega _{\lambda
0}=0.72,$ ${\cal M}=23.94\}$ &  &  \\ 
V2 & \cite{Vishwakarma3} & $\{\Omega _{\texttt{\tiny \rm \rm m}0}=0.79,$ $\Omega _{\lambda
0}=1.41,$ ${\cal M}=23.91\}$ &  &  \\ 
V15 & \cite{Vishwakarma2} & $\{\Omega _{\texttt{\tiny \rm \rm m}0}=0.33,$ $\Omega _{\lambda
0}=0\}$ & 1997ff included &  \\
W1 & \cite{Wang} & $\{H_{0}=65,\Omega _{\texttt{\tiny \rm \rm m}0}=0.7,\Omega _{\lambda
0}=1.2\}$ & Combined data set & \\
\br
\end{tabular}
\item[] $[H_0]={\rm km \, s}^{-1}{\rm Mpc}^{-1}$.
\end{indented}
\end{table}

\begin{table}
\caption{Best-fit parameters other groups (non-standard models).}
\label{tabelle_7}
\begin{indented}
\item[]\begin{tabular}{@{}lllll}
\br
Symbol & Ref. & Best-fit parameters & Comment &  \\ \mr
B1 & \cite{Behnke} & $\{\Omega _{\texttt{\tiny \rm \rm m}0}=0.3,$ $\Omega _{\texttt{\tiny \rm \rm Rigid}
0}=0.7\}$ & Conformal model &  \\ 
V3 & \cite{Vishwakarma3} & $\{\Omega _{\texttt{\tiny \rm \rm m}0}=0.49,$ $\Omega _{\lambda
0}=0.51,$ ${\cal M}=23.97\}$ & $\lambda \sim S^{-2}$ &  \\ 
V4 & \cite{Vishwakarma3} & $\{\Omega _{\texttt{\tiny \rm \rm m}0}=1.86,$ $\Omega _{\lambda
0}=1.52,$ ${\cal M}=23.95\}$ & $\lambda \sim S^{-2}$ &  \\ 
V5 & \cite{Vishwakarma3} & $\{\Omega _{\texttt{\tiny \rm \rm m}0}=0.4,$ $\Omega _{\lambda
0}=0.6,$ ${\cal M}=23.96\}$ & $\lambda \sim H^{2}$ &  \\ 
V6 & \cite{Vishwakarma3} & $\{\Omega _{\texttt{\tiny \rm \rm m}0}=0.98,$ $\Omega _{\lambda
0}=1.53,$ ${\cal M}=23.91\}$ & $\lambda \sim H^{2}$ &  \\ 
V7 & \cite{Vishwakarma3} & $\{\Omega _{\texttt{\tiny \rm \rm m}0}=0.4,$ $\Omega _{\lambda
0}=0.6,$ ${\cal M}=23.96\}$ & $\lambda \sim \mu $ &  \\ 
V8 & \cite{Vishwakarma3} & $\{\Omega _{\texttt{\tiny \rm \rm m}0}=1.62,$ $\Omega _{\lambda
0}=1.59,$ ${\cal M}=23.93\}$ & $\lambda \sim \mu $ &  \\ 
V9 & \cite{Vishwakarma} & $\{\Omega _{\texttt{\tiny \rm \rm m}0}=0.54,$ $\Omega _{\lambda
0}=0.46,$ ${\cal M}=24.03\}$ & Variable $\lambda $ &  \\ 
V10 & \cite{Vishwakarma} & $\{\Omega _{\texttt{\tiny \rm \rm m}0}=1.76,$ $\Omega _{\lambda
0}=1.34,$ ${\cal M}=24.03\}$ & Variable $\lambda $ &  \\ 
V11 & \cite{Vishwakarma2} & $\{\Omega _{\texttt{\tiny \rm \rm m}0}=0.79,$ $\Omega _{\phi
0}=1.41,w_{\phi }=-1\}$ & Quintessence model &  \\ 
V12 & \cite{Vishwakarma2} & $\{\Omega _{\texttt{\tiny \rm \rm m}0}=0.65,$ $\Omega _{\phi
0}=1.22,w_{\phi }=-1\}$ & Quintess. model +1997ff &  \\ 
V13 & \cite{Vishwakarma2} & $\{\Omega _{\texttt{\tiny \rm \rm m}0}=0.52,$ $\Omega _{\lambda
0}=0.48\}$ & $\lambda \sim H^{2}$ +1997ff &  \\ 
V14 & \cite{Vishwakarma2} & $\{\Omega _{\texttt{\tiny \rm \rm m}0}=0.6,$ $\Omega _{\lambda
0}=0.4\}$ & $\lambda \sim S^{-2}$ +1997ff &  \\ 
V16 & \cite{Vishwakarma2} & $\{\Omega _{\lambda 0}=-0.358,$ $z_{\texttt{\tiny \rm \rm max}
}=5\}$ & QSSC model &  \\ 
T1 & \cite{Tomita} & $\{z_{1}=0.08,H_{0}^{II}/H_{0}^{I}=0.87,\Omega
_{0}^{I}=0.3,$ & Model with local void &  \\ 
&  &$H_{0}^{I}=64,\Omega _{0}^{II}=0.6,\lambda _{0}^{II}=0.3\}$   &  &  \\ 
\br
\end{tabular}
\item[] $[H_0]={\rm km \, s}^{-1}{\rm Mpc}^{-1}$.
\end{indented}
\end{table}

\section{Conclusions}\label{CONCLUSION_section}

\begin{figure}[ht]
\setlength{\unitlength}{1mm}
\begin{picture}(80,50)
\epsfig{file=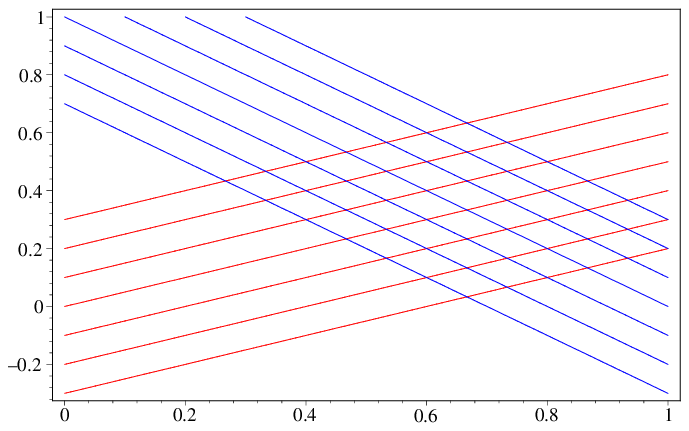}
\put(-62,26){\tiny{$q_0$}}
\put(-62,23){\tiny{-0.3}}
\put(-62,19){\tiny{-0.2}}
\put(-62,16){\tiny{-0.1}}
\put(-62,13){\tiny{0}}
\put(-62,10){\tiny{0.1}}
\put(-62,7){\tiny{0.2}}
\put(-62,4){\tiny{0.3}}
\put(-5,26){\tiny{$\Omega_{k 0}$}}
\put(-5,22){\tiny{-0.3}}
\put(-5,19){\tiny{-0.2}}
\put(-5,16){\tiny{-0.1}}
\put(-5,14){\tiny{0}}
\put(-5,10){\tiny{0.1}}
\put(-5,7){\tiny{0.2}}
\put(-5,4){\tiny{0.3}}
\put(-73,21){\tiny{$\Omega_{\lambda 0}$}}
\put(-35,-1){\tiny{$\Omega_{{\rm m} 0}$}}
\end{picture}
\begin{picture}(80,50)
\epsfig{file=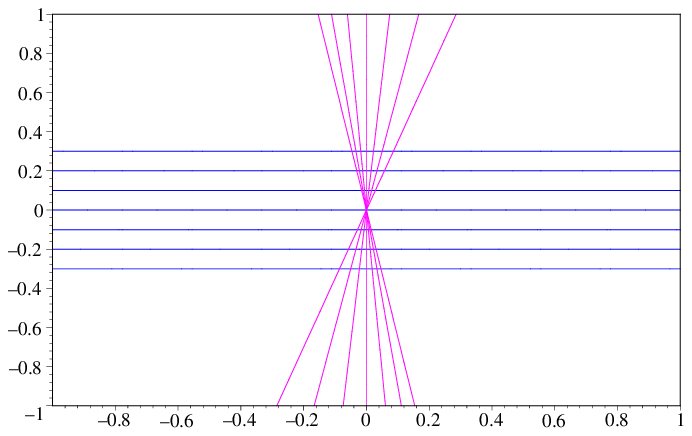}
\put(-22,36){\tiny{$w=-5 \quad \chi=1$}}
\put(-52,30){\tiny{$q_0= -0.3 \, -0.2 \, -0.1 \, 0 \, 0.1 \, 0.2 \, 0.3$}}
\put(-52,13){\tiny{$q_0= 0.3 \, 0.2 \, 0.1 \, 0 \, -0.1 \, -0.2 \, -0.3$}}
\put(-73,21){\tiny{$\Omega_{k 0}$}}
\put(-35,-1){\tiny{$\Omega_{\zeta 0}$}}
\end{picture}
\begin{picture}(80,50)
\epsfig{file=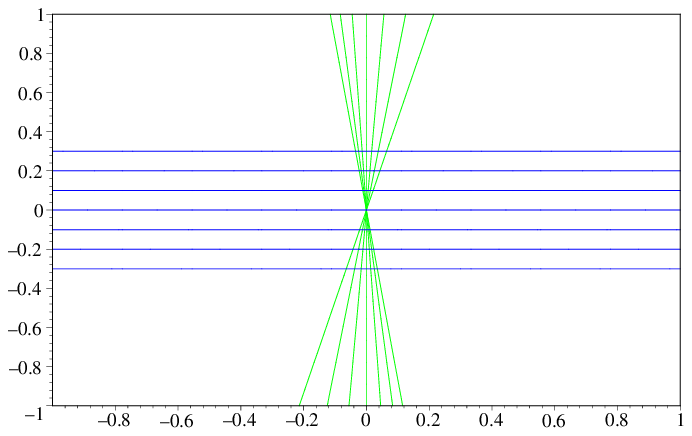}
\put(-22,36){\tiny{$w=-1 \quad \chi=1$}}
\put(-52,30){\tiny{$q_0= -0.3 \, -0.2 \, -0.1 \, 0 \, 0.1 \, 0.2 \, 0.3$}}
\put(-52,13){\tiny{$q_0= 0.3 \, 0.2 \, 0.1 \, 0 \, -0.1 \, -0.2 \, -0.3$}}
\put(-73,21){\tiny{$\Omega_{k 0}$}}
\put(-35,-1){\tiny{$\Omega_{\zeta 0}$}}
\end{picture}
\begin{picture}(80,50)
\epsfig{file=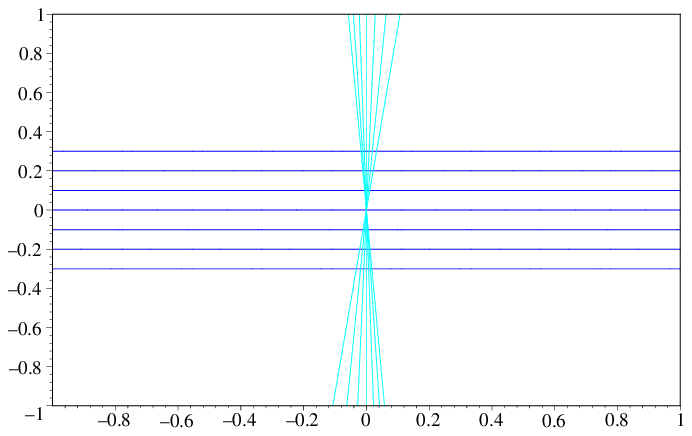}
\put(-22,36){\tiny{$w=0 \quad \chi=1$}}
\put(-52,30){\tiny{$q_0= -0.3 \, -0.2 \, -0.1 \, 0 \, 0.1 \, 0.2 \, 0.3$}}
\put(-52,13){\tiny{$q_0= 0.3 \, 0.2 \, 0.1 \, 0 \, -0.1 \, -0.2 \, -0.3$}}
\put(-73,21){\tiny{$\Omega_{k 0}$}}
\put(-35,-1){\tiny{$\Omega_{\zeta 0}$}}
\end{picture}
\begin{picture}(80,50)
\epsfig{file=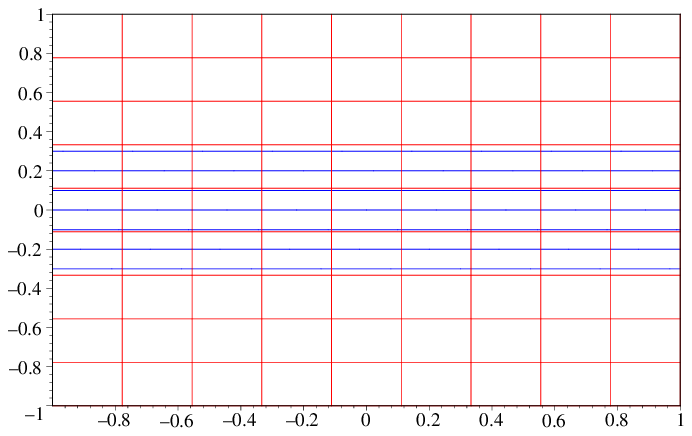}
\put(-22,34){\tiny{$w=1 \quad \chi=1$}}
\put(-56,8){\tiny{$q_0=0$}}
\put(-73,21){\tiny{$\Omega_{k 0}$}}
\put(-35,-1){\tiny{$\Omega_{\zeta 0}$}}
\end{picture}
\begin{picture}(80,50)
\epsfig{file=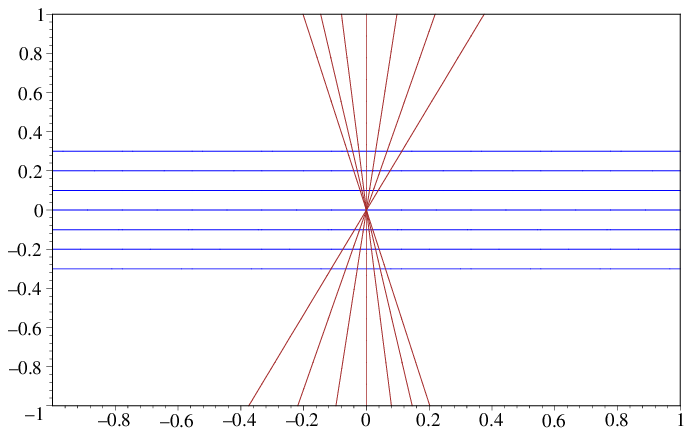}
\put(-22,36){\tiny{$w=5 \quad \chi=1$}}
\put(-52,30){\tiny{$q_0= -0.3 \, -0.2 \, -0.1 \, 0 \, 0.1 \, 0.2 \, 0.3$}}
\put(-52,13){\tiny{$q_0= 0.3 \, 0.2 \, 0.1 \, 0 \, -0.1 \, -0.2 \, -0.3$}}
\put(-73,21){\tiny{$\Omega_{k 0}$}}
\put(-35,-1){\tiny{$\Omega_{\zeta 0}$}}
\end{picture}
\caption[Contours of the deceleration factor]{Contour lines of the
  deceleration factor $q_0$ and the density parameter $\Omega_{k0}$ in the
  ($\Omega_{\lambda 0},\Omega_{{\rm m}0}$) and ($\Omega_{k 0},\Omega_{\zeta
  0}$) plane, respectively. The figure on the top left corresponds to the
  standard model which contains only usual matter $w=0$ and a contribution
  from the cosmological constant. The other figures belong to the Weyl-Cartan
  model in case of different choices of the equation of state parameter $w$ and a vanishing induced cosmological constant.}
\label{figure_3}
\end{figure}

\begin{figure}[ht]
\setlength{\unitlength}{1mm}
\begin{picture}(80,50)
\epsfig{file=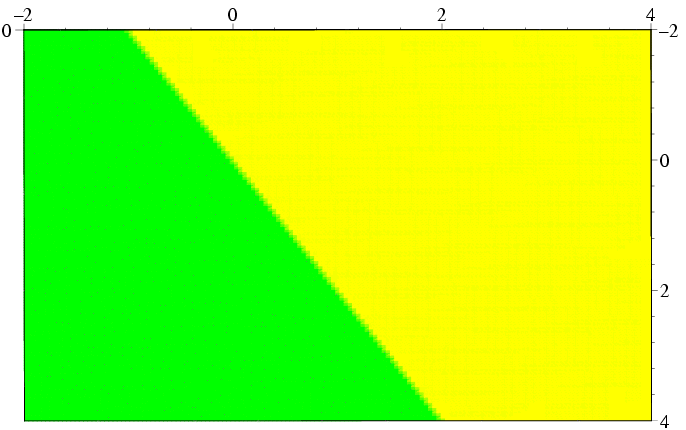}
\put(-73,21){\tiny{$\Omega_{{\rm m} 0}$}}
\put(-35,-2){\tiny{$\Omega_{\lambda 0}$}}
\put(-28,27){\tiny{accelerating}}
\put(-60,12){\tiny{decelerating}}
\end{picture}
\begin{picture}(80,50)
\epsfig{file=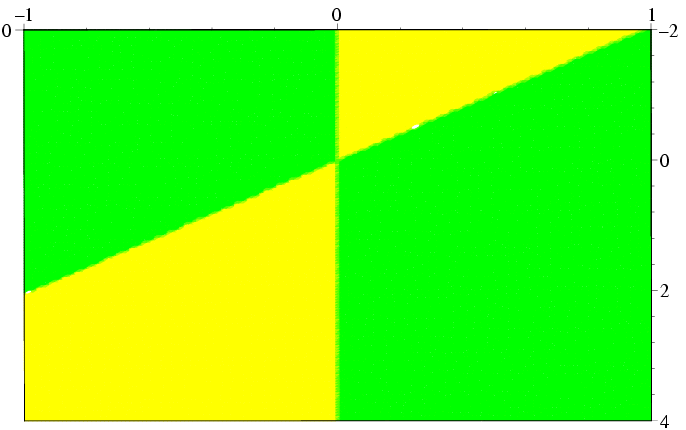}
\put(-72,21){\tiny{$\Omega_{k 0}$}}
\put(-35,-2){\tiny{$\Omega_{\zeta 0}$}}
\put(-25,6){\tiny{$w=-0.9,\, \chi=1$}}
\put(-55,12){\tiny{accelerating}}
\put(-28,20){\tiny{decelerating}}
\put(-33,36){\tiny{accelerating}}
\put(-60,30){\tiny{decelerating}}
\end{picture}
\caption[]{Sign of the deceleration factor $q_0$ in the density parameter plane for the standard FLRW model (lhs) and for the Weyl-Cartan model (rhs).}
\label{figure_5}
\end{figure}

\paragraph{Fitting results}
As we have shown in the previous section it is possible to describe the
observational data within both models. It is noteworthy that without any
additional constraints the best-fit parameters, i.e.\ set F13 in table
\ref{tabelle_4}, within the FLRW model favour a closed universe, whereas the
best-fit solution within the Weyl-Cartan model, i.e.\ set C2 in table
\ref{tabelle_5}, corresponds to an open universe. In case of the FLRW model 
our results comply with the ones of Wang found in \cite{Wang}.
If one wants to impose the condition of spatial flatness, the best-fit
parameters are given by $({\rm F28},65,0.29,0.71,135.26,1.51,-0.56)$ within the FLRW model,
and $({\rm C9},69,0,-0.28,1,-0.5,292.821,3.32,1)$ within the Weyl-Cartan
model\footnote{Note that here we make use of the same enumeration as in table
  \ref{tabelle_4} and table \ref{tabelle_5}.}. This is an interesting result
since as soon as we assume that the universe is flat we are not able to find
parameters within our new model which fit the data. This is in
contrast to the FLRW model where the assumption of spatial flatness worsens
the fit only slightly. Hence, at least within the parameter intervals we
considered, our model does not support a flat universe.    
Note that the fit F28 is in compliance with the current cosmological 
concordance model which encompasses about 30\% of matter and a dark energy contribution of about 70\%.  
\paragraph{Deceleration factor}
The best-fit sets F13 and C2 in both models predict a universe which
is presently in an accelerating phase of expansion. In case of the Weyl-Cartan
model the current value of the deceleration parameter is roughly seven times
smaller than in the FLRW case, hence in our model the expansion of the
universe seems to accelerate less rapid than predicted by the standard
scenario. Figure \ref{figure_5} provides an overview over the sign of the deceleration
factor in the FLRW and in the Weyl-Cartan model. In case of the latter we plotted the
distribution in the parameter plane which contains the best-fit C2. Another interesting 
property associated with deceleration factor is
the fact that it seems to provide an independent cosmological test. As
displayed in the plot on top lhs in figure \ref{figure_3} the contour lines
for a constant deceleration factor within the $(\Omega_{{\rm m}0},\Omega_{\lambda 0})$-parameter plane of the FLRW model intersect the ones for constant
$\Omega_{k0}$ at a non-zero angle. Hence if we assume that we can measure
$\Omega_{k0}$ via analysis of the fluctuations within the CMB \cite{LiddleLyth} then we are able
to pin down the pair $(\Omega_{{\rm m}0},\Omega_{\lambda 0})$. The situation
within the new model is similar to the one FLRW case, i.e.\ the curves of
constant $q_0$ and $\Omega_{k0}$ intersect each other at a non-zero
angle. Figure \ref{figure_3} contains plots for five different choices of the
equation of state parameter $w$. As we can see from plot at the bottom lhs
only the choice $w=1$ corresponds to a degenerated situation, since in this
case the deceleration factor vanishes according to (\ref{EXTWEYL_present_day_deceleration_factor_in_the_lambda_zero_model}).     

\begin{figure}[ht]
\setlength{\unitlength}{1mm}
\begin{picture}(80,50)
\epsfig{file=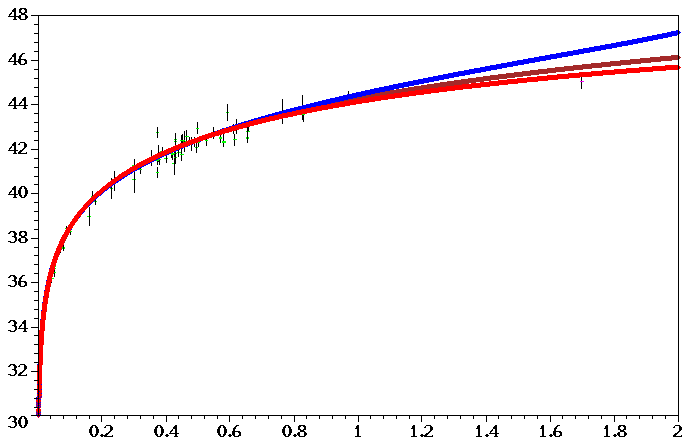}
\put(-78,24){\tiny{$\mu$}}
\put(-40,1){\tiny{$z$}}
\put(-40,40){\tiny{C2}}
\put(-40,30){\tiny{F13}}
\put(-40,34){\tiny{F28}}
\end{picture}
\begin{picture}(80,50)
\epsfig{file=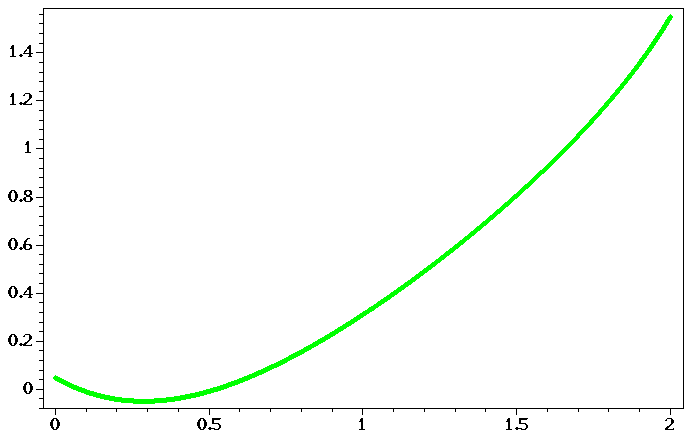}
\put(-80,24){\tiny{$\Delta\mu$}}
\put(-40,1){\tiny{$z$}}
\put(-50,30){\tiny{$\mu_{_{\rm C2}}-\mu_{_{\rm F13}}$}}
\end{picture}
\caption[]{On the lhs we plotted the distance modulus versus the redshift for
  the best-fit models F13, F28, and C2 up to $z=2$. As one can see from the rhs the
  difference between both models grows with the redshift.}
\label{figure_4}
\end{figure}

\paragraph{Other groups}
As we have shown in sections \ref{Mag_redshit_in_standard_model} and
\ref{EXTWEYL_magnitude_redshift_relation} the magnitude-redshift relation
depends on several assumptions. Especially its strong dependence on the
underlying field equations renders it to be a useful tool to discriminate
between different cosmological models. In table \ref{tabelle_6} and \ref{tabelle_7}
we collected some of the results from several other groups who used this
relation within the FLRW as well as in non-standard models. It becomes clear
from table \ref{tabelle_6} that in addition to the model dependence of the
magnitude-redshift relation, the estimates for the cosmological parameters
strongly depend on the data set which is used to perform a fit. It is
interesting that some of the early best-fit parameter sets like S1 and
G1, e.g., correspond to unphysical models. The situation within non-standard
models is similar. As one can infer from table \ref{tabelle_7} the parameters
estimated within non-standard scenarios depend strongly on the underlying
model. Although it seems to be possible to describe the supernova data equally
well within several alternative scenarios, the main benefit of the
cosmological standard model consists of its level of detail, which results in
the availability of a number of different cosmological tests \cite{Peebles1,Lineweaver,Jaffe}. Note that without the constraints from other cosmological tests the used data set does not favour a flat universe. The best-fit F13 is even slightly better than the one which is obtained if we impose the constraint of spatial flatness right from the beginning, i.e.\ the fit F28 mentioned at the beginning of this section. It is noteworthy that nearly all groups come to the same result that a non-vanishing cosmological constant seems to be inevitable for the description of the supernova data within FLRW model, as becomes clear from table \ref{tabelle_6} and our results in table \ref{tabelle_4}.

\paragraph{Summary \& Outlook }

In figure \ref{figure_4} we plotted the distance modulus versus the redshift
up to $z=2$. The upper curve corresponds to the best-fit within the
Weyl-Cartan model. It becomes clear that the supernovae at high redshifts will
appear dimmer within this model than in the FLRW case. The data point at $z=1.7$ corresponds to the farthest known
supernova 1997ff as reported by Riess \etal \cite{Riess1}. Although not
significant this supernova seems to favour the best-fit FLRW model F13. The uncertainties connected with 1997ff are large, therefore we did not
include it in our fitting procedure. Note that also the best-fit flat model
F28 seems not to fit the data from 1997ff. A possible magnification of 1997ff by gravitational lensing was discussed by M\"ortsell \etal in \cite{Moertsell1}. It is too early to make a reliable
prediction at this point, one has to wait until more data at high
redshifts becomes available. As we can see from the rhs in figure
\ref{figure_4} a survey at high redshifts (like SNAP e.g.) should enable us to
discriminate between the two best-fit models C2 and F13.

It is also possible to derive the age of the universe within our model. In case
of the parameter set C2 we obtain an age of $18.4$ Gyr. Thus, our
model combined with the supernova data yields an older universe than the
standard model - i.e.\ $13.8$ Gyr in case of F13, and $14.6$ Gyr for the flat
scenario F28. Hence, there is {\it no} conflict with the estimates from the
observations of globular clusters and nucleocosmochronology which range from $11$ to $15$ Gyr \cite{Chaboyer, Truran}.

We reviewed the derivation of the magnitude-redshift relation within the
cosmological standard model with special emphasis on the various assumptions
which are necessary to obtain this relation. We have shown that it is possible to find a similar relation between the luminosity and distance within the new
Weyl-Cartan model proposed in \cite{Extweyl}. We performed fits to the
combined data set of Wang \cite{Wang} within both models. Note that this is
the first time that a fit of MAG based model to a real data set has been
performed. Our efforts can be viewed as a first step to relief the Weyl-Cartan
model of its current toy model character. We are aware that at the moment our model
is far from being called {\it realistic}. Nevertheless we were able to show that 
in principle it is possible to obtain a magnitude-redshift relation within our 
MAG based model. Despite this success it is too early to use the best-fit parameters 
obtained in this work in order to pin down the free parameters within the model. 
One needs at least another independent cosmological test in order to impose meaningful
constraints. Therefore, our aim in the next article of this series is to
check whether our model is compatible with recent observations of the cosmic
microwave background and if there exists a concordance region with the parameters 
obtained in this work.  

\ack
The author is grateful to F.W.\ Hehl, G.\ Rubilar, and the members of the
gravity group at the University of Cologne for their support. Special thanks
go to Y.\ Wang for valuable comments and for providing her SNe data set. \bigskip

\appendix

\section{Units\label{NATURAL_UNITS}}

In this work we made use of \textit{natural units}, i.e. $\hbar =c=1$
(cf table \ref{tabelle_8}).
\begin{table}
\caption{Natural units.}
\label{tabelle_8}
\begin{indented}
\item[]\begin{tabular}{@{}llll}
\br
[energy] & [mass] & [time] & [length] \\ 
\mr
length$^{-1}$ & length$^{-1}$ & length & length \\ 
\br
\end{tabular}
\end{indented}
\end{table}
Additionally, we have to be careful with the coupling constants and the
coordinates within the coframe. In order to keep things as clear as
possible, we provide a list of the quantities emerging throughout all sections in table \ref{tabelle_9}.
\begin{table}
\caption{Units of quantities.}
\label{tabelle_9}
\begin{indented}
\item[]\begin{tabular}{@{}lll}
\br
Quantities & \textit{1} & \textit{Length} \\ \mr
Gauge potentials & $[g_{\alpha \beta }]$& $[\vartheta ^{\alpha }]$ \\ 
Field strengths & $[Q_{\alpha \beta }],[\tilde{R}_{\alpha \beta }]$ &
\\ 
Matter currents &  & $[\Sigma _{\alpha }]^{-1}$ \\ \mr
Coordinates & $[\theta ],[\phi ],[r]$ & $[t]$ \\\mr 
Functions & $[\Omega_w],[\Omega_k],[\Omega_{\rm m}],[\Omega_\lambda],$ & $[S],[\mu]^{-\frac{1}{4}},[p]^{-\frac{1}{4}},$ \\ 
& $[\zeta],[\Omega_{\rm total}],[\Omega_\zeta],[\Omega_\Lambda]$ &
$[p_{r}]^{-\frac{1}{4}},[H]^{-1}$ \\\mr
Miscellany & $[u^\alpha],[z],[q],[w],$ & $[\Sigma _{\alpha \beta }]^{-\frac{1}{4}},[d_{\tiny \rm luminosity}],$
\\
&$[m],[M],[{\cal M}],[\mu]$&$[\breve{L}]^{-2},[\breve{F}]^{-4}$ \\\mr 
Constants & $[\chi ],[b],[k],[a_{I}]$ & $[\kappa ]^{\frac{1}{2}},[\lambda]^{-\frac{1}{2}},[\Lambda]^{-\frac{1}{2}},[c]^{-\frac{1}{2}},$ \\ 
&$$  & $[\Xi ]^{-\frac{1}{4}},[G]^{\frac{1}{2}},[\varkappa_1]^{\frac{1}{-4+3(1+w)}}$ \\ \br
\end{tabular}
\end{indented}
\end{table}
Note that $[d]=1$ and $[\,^{\star }]=$ length$^{n-2p}$, where $n=$ dimension
of the spacetime, $p=$ degree of the differential form on which $^{\star }$
acts.

\end{document}